\documentclass[11pt,a4paper]{article}
\usepackage{authblk}
\usepackage{amsmath,amsfonts,amssymb}
\usepackage{bbold}
\usepackage{bbm}
\usepackage{txfonts}
\usepackage{tensor}
\usepackage[T1]{fontenc}
\usepackage{hyperref}

\newtheorem{proof}{Proof}

\numberwithin{equation}{section}
\numberwithin{theorem}{section}
\numberwithin{proposition}{section}
\numberwithin{remark}{section}
\numberwithin{example}{section}

\def\PL#1{\textit{Phys.\ Lett.}\ {\bf#1}}\def\CMP#1{\textit{Commun.\ Math.\ Phys.}\ {\bf#1}}

\def\PR#1{\textit{Phys.\ Rev.}\ {\bf#1}}\def\CQG#1{\textit{Class.\ Quantum Grav.}\ {\bf#1}}
\def\NP#1{\textit{Nucl.\ Phys.}\ {\bf#1}}
\def\JMP#1{\textit{J.\ Math.\ Phys.}\ {\bf#1}}
\def\PRS#1{\textit{Proc.\ R. Soc.\ Lond.}\ {\bf#1}}
\def\JoP#1{\textit{J.\ Phys.}\ {\bf#1}} \def\IJMP#1{\textit{Int.\ J. Mod.\ Phys.}\ {\bf #1}}

\def\JHEP#1{\textit{JHEP}\ {\bf#1}}\def\JCAP#1{\textit{JCAP}\ {\bf#1}}

\def\arx#1{{\tt arXiv:#1}}
\def\EPJ#1{\textit{Eur.\ Phys.\ J.}\ {\bf#1}}

\newcommand{\bibp}[3]{#1, #2, #3}

\title{Dual $\kappa$-Minkowski spaces and $\kappa$-Poincar\'{e} algebras from Yang model and their Weyl realizations}

\begin{document}
\author[a]{Tea Martini\'{c} Bila\'{c}}
\affil[a]{Faculty of Science, University of Split, Rudjera Bo\v{s}kovi\'{c}a 33, 21000 Split, Croatia\\
	E-mail:\;\href{mailto:teamar@pmfst.hr}{teamar@pmfst.hr}}

\author[b]{Stjepan Meljanac}
\affil[b]{Rudjer Bo\v{s}kovi\'{c} Institute, Theoretical Physics Division, Bijeni\v{c}ka c. 54, HR 10002 Zagreb, Croatia\\
	E-mail:\;\href{mailto:meljanac@irb.hr}{meljanac@irb.hr}}

 \author[c]{Salvatore Mignemi}
 \affil[c]{Dipartimento di Matematica, Universit\`{a} di Cagliari via Ospedale 72, 09124 Cagliari, Italy and INFN, Sezione di Cagliari Cittadella Universitaria, 09042 Monserrato, Italy\\
 	E-mail:\;\href{mailto:smignemi@unica.it}{smignemi@unica.it}}

\date{}
\maketitle

\begin{abstract}
	
	We consider the Yang algebras isomorphic to $o(1,5), o(2,4), o(3,3)$  and derive dual $\kappa$-Minkowski and $\kappa$-Poincar\'{e} algebras in terms of a metric $g$. The corresponding Weyl realization is presented and coproduct, star product and twist are computed in terms of the metric $g$. Finally, we construct reduced  $\kappa$-Minkowski and $\kappa$-Poincar\'{e} algebras as special cases.
\end{abstract}

\section{Introduction}

It is well known that the principles of quantum theory and general relativity are incompatible
and a more general theory reconciling them is necessary. This will presumably be based on new
assumptions on the foundations of the two theories.
A debated possibility is that the quantum phase space must be modified, for example deforming the
canonical Heisenberg relations. This of course would induce observable effects, as a generalized
uncertainty principle (GUP) \cite{gup}.
However, it is evident on dimensional grounds that such effects are related to the Planck
scale and therefore are besides the present experimental reach.

By definition, a deformation of the commutation relations of position operators  induces a
noncommutativity of spacetime \cite{DF} that can be related to the curvature of momentum space
\cite{KG}.
On the other hand, it is well known that the curvature of
spacetime gives rise to a deformation of the commutation relations of momentum operators.
In general, both deformations could be considered simultaneously.

This idea was first advanced by C.~N.~Yang in \cite{Ya}, who, building on an earlier
proposal of Snyder \cite{Sn}, proposed an algebra isomorphic to the
fifteen-dimensional orthogonal algebra $o(1,5)$, consisting in a combination of the Heisenberg
algebra of a deformed phase space with noncommuting coordinates $\hat x_\mu$ and $\hat  p_\mu$
and the Lorentz algebra generated by $M_{\mu\nu}$, together with a further
scalar generator $\hat h$, necessary for the closure.
The Yang algebra depends on two deformation parameters $M$ and $R$ that fix the scale of
the curvature of momentum and position spaces, and are often identified with
the Planck mass and the de Sitter radius.
The Yang model enjoys the remarkable property of being invariant under a generalized Born duality
\cite{Bo}, which interchanges positions and momenta.
Moreover, taking suitable limits, it can be reduced to the Snyder algebra or its dual de Sitter algebra.

After some time, the study of the Yang model was resumed in recent years. Some generalizations were presented by Khruschev and Leznov in \cite{KL}, while in ref. \cite{Meljanac-1} an extension of the model that includes also the related triply special relativity (TSR) theory \cite{Kowalski-1, Mignemi-1} was introduced. Further investigations concerning in particular its realizations on a canonical phase space have been recently performed in \cite{MM-1}-\cite{Lukierski-1}, using the methods introduced in \cite{Battisti-1}-\cite{Girelli} for the study of Snyder space. Other contributions to the study of the Yang model are given in \cite{Guo}-\cite{Lukierski-2}, while different models of noncommutative geometry in curved spaces can be found in \cite{Mignemi-2}-\cite{Aschieri}. Also, relations of the Yang model with other theories, as for example conformal field theory or fuzzy gravity, have been investigated in \cite{others}.

Using the methods presented in \cite{Battisti-1, MM-3} for Snyder space, in \cite{MS-1}-\cite{Lukierski-3} for $\kappa$-deformed Snyder spaces, and in \cite{Meljanac-3} for isomorphism between the Yang model and orthogonal algebras, the Hopf algebra structure,  coproduct, twist and the associative star product were obtained for the Yang model \cite{MMM}.

Recently in \cite{LMP1, MMM}, it was shown that the Yang algebra can be subjected to a further deformation of
$\kappa$ type \cite{kappa}, depending on two different $\kappa$ parameters.  This doubly $\kappa$-deformed Yang model is still isomorphic to the $o(1,5)$ algebra.
This proposal was investigated in more detail in \cite{MMM} where the change of basis necessary to pass
from the original Yang algebra to the new one was explicitly worked out,
and Weyl realization of such algebra in extended phase space were given. Later, in \cite{LMP2} a new relativistic doubly
 $\kappa$-deformed quantum phase space was considered.
In this way, a generalized $\kappa$-Poincar\'e algebra is obtained, that displays a duality of Born type
between position and momentum deformations.
We call the associated spacetimes dual $\kappa$-Minkowski spaces.

In this paper, we analyize the properties of the doubly $\kappa$-Poincar\'e algebra and the
associated $\kappa$-Minkowski space in a
way analogous to that used in ref.~\cite{MMM} for the doubly $\kappa$-deformed Yang models for the cases of $o(1,5)$, $o(2,4)$ and $o(3,3)$ algebras. The results are written in terms of a metric $g$; dual $\kappa$-Minkowski and dual $\kappa$-Poincar\'e algebras are obtained as special cases for $g_{44} = g_{55} = 0$. Weyl realization, coproduct, star product and twist are presented in terms of the metric $g$. Finally, reduced dual $\kappa$-Minkowski and dual $\kappa$-Poincar\'e algebras, where the scalar generator $\hat h$ is taken as dependent \cite{MM-5}, are constructed.

 \section{Generalized Yang models isomorphic to $o(1,5,g)$}
 \label {s-2}
There are infinitely many Born-dual $\kappa$-Minkowski spaces among non linear algebras satisfying the Jacobi Identities \cite{LMP2}.
In this section we review and generalize the construction of a class of Lie algebra-type spaces containing dual $\kappa$-Minkowski spaces that can be obtained from the Yang model \cite{MMM}
with a suitable choice of the parameters.

Let us start with the Yang algebra \cite{Ya}
\begin{equation}\label{0.1}
\left[  \hat{x}_{\mu},\hat{x}_{\nu}\right]  =\frac{i\epsilon_{1}}{M^{2}}M_{\mu\nu}, \quad \left[  \hat{p}_{\mu},\hat{p}_{\nu}\right] =\frac{i\epsilon_{2}}{R^{2}}M_{\mu\nu},
\end{equation}
\begin{equation}\label{0.2}
\left[  M_{\mu\nu}, \hat{x}_{\lambda}\right]   =i\left( \eta_{\mu\lambda}\hat{x}_{\nu}-\eta_{\nu\lambda}\hat{x}_{\mu}\right) ,\quad  \left[  M_{\mu\nu}, \hat{p}_{\lambda}\right]   =i\left( \eta_{\mu\lambda}\hat{p}_{\nu}-\eta_{\nu\lambda}\hat{p}_{\mu}\right),
\end{equation}
\begin{equation}\label{0.3}
\left[  \hat{x}_{\mu}, \hat{p}_{\nu}\right] =i\eta_{\mu\nu}\hat{h}, \quad \left[  \hat{h}, \hat{x}_{\mu}\right]  =\frac{i\epsilon_{1}}{M^{2}}\hat{p}_{\mu}, \quad \left[  \hat{h}, \hat{p}_{\mu}\right]  =-\frac{i\epsilon_{2}}{R^{2}}\hat{x}_{\mu},
	\end{equation}
\begin{equation}\label{0.4}
\left[ M_{\mu\nu}, \hat{h} \right]  =0,
\end{equation}
\begin{equation}\label{0.5}
\left[  M_{\mu\nu},M_{\rho\sigma}\right]  =i\left( \eta_{\mu\rho}M_{\nu\sigma}-\eta_{\mu\sigma}M_{\nu\rho}-\eta_{\nu\rho}M_{\mu\sigma}+\eta_{\nu\sigma}M_{\mu\rho}\right),
\end{equation}
where $\epsilon_{1}^{2}=\epsilon_{2}^{2}=1$ and $\;\eta_{\mu\nu}=diag(-1,1,1,1)$.

In this section we shall consider the case $\epsilon_{1} = \epsilon_{2} =1$  corresponding to the $o(1,5)$ algebra.
\\

Let us define the most general new generators linear in $\hat{x}_{\mu},\; \hat{p}_{\mu}\;$ and $M_{\mu\nu}$ introducing parameters $\kappa$ and $\tilde{\kappa}$ generalizing the results of \cite{MMM},
\begin{equation}\label{x-2}
\tilde{X}_{\mu}=u\left( \cos \varphi \hat{x}_{\mu}+\frac{R}{M}\sin \varphi \hat{p}_{\mu}\right) +\frac{1}{\kappa} a_{\rho}M_{\mu\rho},
\end{equation}
\begin{equation}\label{p-2}
\tilde{P}_{\mu}=v\left( \cos \psi \hat{p}_{\mu}+\frac{M}{R}\sin \psi \hat{x}_{\mu}\right) +\frac{1}{\tilde{\kappa}} b_{\rho}M_{\mu\rho},
\end{equation}
with the Lorentz generators $\tilde{M}_{\mu\nu}=M_{\mu\nu}$ unchanged. The scalar parameters $u,v,\varphi, \psi$ and the four-vectors $a_{\mu},b_{\mu}\;$ are dimensionless  with $uv\neq 0$.\\
The inverse transformations are
\begin{equation}\label{0.6}
\hat{x}_{\mu}=\frac{u^{-1}\frac{1}{R}\cos \psi\left(\tilde{X}_{\mu}-\frac{1}{\kappa} a_{\rho}M_{\mu\rho}\right) -v^{-1}\frac{1}{M}\sin \varphi\left(\tilde{P}_{\mu}-\frac{1}{\tilde{\kappa}} b_{\rho}M_{\mu\rho}\right) }{\frac{1}{R} \cos \left( \varphi +\psi\right) },
\end{equation}
\begin{equation}\label{0.7}
\hat{p}_{\mu}=\frac{v^{-1}\frac{1}{M}\cos \varphi\left(\tilde{P}_{\mu}-\frac{1}{\tilde{\kappa}} b_{\rho}M_{\mu\rho}\right)-u^{-1}\frac{1}{R}\sin \psi\left(\tilde{X}_{\mu}-\frac{1}{\kappa} a_{\rho}M_{\mu\rho}\right)}{\frac{1}{M} \cos \left( \varphi +\psi\right)}
.\end{equation}

The generators $\tilde{X}_{\mu}, \;\tilde{P}_{\mu},\; M_{\mu\nu}$ and $\tilde{H}$ generate a new class of Lie algebras isomorphic to the initial Yang algebra. These algebras are defined by
\begin{equation}\label{1.1}
\left[  \tilde{X}_{\mu}, \tilde{X}_{\nu} \right]  =i\left( \frac{1}{M^{2}}\left(u^{2} +a^{2}\frac{M^{2}}{\kappa^{2}}\right) M_{\mu\nu}+\frac{1}{\kappa}\left( a_{\mu}\tilde{X}_{\nu}-a_{\nu}\tilde{X}_{\mu} \right)\right) ,
\end{equation}
\begin{equation}\label{1.2}
\left[  \tilde{P}_{\mu}, \tilde{P}_{\nu}\right]  =i\left( \frac{1}{R^{2}}\left(v^{2} +b^{2}\frac{R^{2}}{\tilde{\kappa}^{2}}\right)M_{\mu\nu}+\frac{1}{\tilde{\kappa}}\left( b_{\mu}\tilde{P}_{\nu}-b_{\nu}\tilde{P}_{\mu}\right)\right) ,
\end{equation}
\begin{equation}\label{1.3}
\left[  \tilde{X}_{\mu}, \tilde{P}_{\nu}\right] =i\left( \eta_{\mu\nu}\tilde{H}+\frac{1}{\tilde{\kappa}} b_{\mu}\tilde{X}_{\nu}-\frac{1}{\kappa}a_{\nu}\tilde{P}_{\mu}+\frac{1}{MR} \left( uv \sin(\varphi+\psi)+abMR\frac{1}{\kappa\tilde{\kappa}}\right) M_{\mu\nu}\right) ,
\end{equation}
\begin{equation}\label{1.4}
\left[  M_{\mu\nu},\tilde{X}_{\lambda}\right]  =i\left( \eta_{\mu\lambda}\tilde{X}_{\nu}-\eta_{\nu\lambda}\tilde{X}_{\mu}+\frac{1}{\kappa}\left( a_{\mu}M_{\lambda\nu}-a_{\nu}M_{\lambda\mu}\right) \right),
\end{equation}
\begin{equation}\label{1.5}
\left[  M_{\mu\nu},\tilde{P}_{\lambda}\right]  =i\left( \eta_{\mu\lambda}\tilde{P}_{\nu}-\eta_{\nu\lambda}\tilde{P}_{\mu}+\frac{1}{\tilde{\kappa}}\left( b_{\mu}M_{\lambda\nu}-b_{\nu}M_{\lambda\mu}\right) \right),
\end{equation}
where
\begin{equation}\label{1.6}
\tilde{H}=\hat{h}uv\cos(\varphi+\psi)+\frac{1}{\kappa} a\tilde{P}-\frac{1}{\tilde{\kappa}} b \tilde{X}-\frac{1}{\kappa\tilde{\kappa}} a_{\rho}b_{\sigma}M_{\rho\sigma}
\end{equation}
and
\begin{equation}\label{1.7}
\left[  M_{\mu\nu},\tilde{H}\right]  =i\left( \frac{1}{\tilde{\kappa}}\left( b_{\nu}\tilde{X}_{\mu}-b_{\mu}\tilde{X}_{\nu}\right) -\frac{1}{\kappa}\left( a_{\nu}\tilde{P}_{\mu}-a_{\mu}\tilde{P}_{\nu}\right)\right) ,
\end{equation}
\begin{equation}\label{1.8}
\left[  \tilde{H}, \tilde{X}_{\mu}\right]  =i\left(\frac{1}{M^{2}}\left(u^{2} +a^{2}\frac{M^{2}}{\kappa^{2}}\right) \tilde{P}_{\mu}-\frac{1}{MR} \left( uv \sin(\varphi+\psi)+abMR\frac{1}{\kappa\tilde{\kappa}}\right)\tilde{X}_{\mu}-\frac{1}{\kappa} a_{\mu}\tilde{H}\right),
\end{equation}
\begin{equation}\label{1.9}
\left[  \tilde{H}, \tilde{P}_{\mu}\right] =-i\left(\frac{1}{R^{2}}\left(v^{2} +b^{2}\frac{R^{2}}{\tilde{\kappa}^{2}}\right) \tilde{X}_{\mu}-\frac{1}{MR} \left( uv \sin(\varphi+\psi)+abMR\frac{1}{\kappa\tilde{\kappa}}\right)\tilde{P}_{\mu}+\frac{1}{\tilde{\kappa}} b_{\mu}\tilde{H}\right).
\end{equation}

The generalized Born duality \cite{Bo} still holds for
$M \leftrightarrow R, \;\kappa \leftrightarrow \tilde{\kappa},\; a_{\mu}\rightarrow -b_{\mu},\; b_{\mu}\rightarrow a_{\mu},\; u \leftrightarrow v,\; \varphi\rightarrow -\psi,\; \psi \rightarrow -\varphi,\; \tilde{X}_{\mu} \rightarrow  -\tilde{P}_{\mu},\; \tilde{P}_{\mu} \rightarrow \tilde{X}_{\mu},\; M_{\mu\nu}\leftrightarrow M_{\mu\nu},\; \tilde{H} \leftrightarrow \tilde{H},\; \epsilon_{1}\leftrightarrow \epsilon_{2}.$
These commutation relations are similar to those of \cite{MMM,LMP2}.
If $\frac{a_{\mu}}{\kappa}=0\;$ and $\frac{b_{\mu}}{\tilde{\kappa}}=0$ then above algebra reduces to the Khruschev-Leznov model \cite{KL} with $\rho=uv \sin (\varphi+\psi).$
\bigskip

If we define generators
\begin{equation}
M_{\mu 4}=M \hat{x}_{\mu}, \; M_{\mu 5}=R\hat{p}_{\mu}, \;  M_{45}=MR \hat{h}
\end{equation}
then the Yang algebra \eqref{0.1}- \eqref{0.5} takes the form of an orthogonal algebra $o(1,5)$ with
\begin{equation}\label{2}
\left[  M_{A B},M_{C D}\right]  =i\left( \eta_{A C}M_{B D}-\eta_{A D}M_{B C}+\eta_{B D}M_{A C}-\eta_{B C}M_{A D}\right),
\end{equation}
where $A, B, C, D=0,1,2,3,4,5,\; \eta_{AB}=\text{diag}(-1,1,1,1,1,1).$
\bigskip

Furthermore, if we define
\begin{equation}\label{2-d}
\tilde{M}_{\mu 4}=M \tilde{X}_{\mu}, \; \tilde{M}_{\mu 5}=R\tilde{P}_{\mu}, \;  \tilde{M}_{45}=MR \tilde{H}
\end{equation}
then the algebra \eqref{1.1}-\eqref{1.9} becomes the $o(1,5,g)$ algebra with
\begin{equation}\label{2.1}
\left[  \tilde{M}_{A B},\tilde{M}_{C D}\right]  =i\left( g_{A C}\tilde{M}_{B D}-g_{A D}\tilde{M}_{B C}+g_{B D}\tilde{M}_{A C}-g_{B C}\tilde{M}_{A D}\right),
\end{equation}
where the metric $g_{AB}$ is given by a symmetric matrix defined as \begin{align}
&g_{\mu\nu}=\eta_{\mu\nu}, \; g_{\mu 4}=\frac{M}{\kappa}a_{\mu}, \; g_{\mu 5}=\frac{R}{\tilde{\kappa}}b_{\mu}, \; g_{44}=u^{2} +a^{2}\frac{M^{2}}{\kappa^{2}}, \; g_{55}=v^{2} +b^{2}\frac{R^{2}}{\tilde{\kappa}^{2}},\;\notag  \\
& g_{45}=g_{54}= uv \sin(\varphi+\psi)+ab\frac{MR}{\kappa\tilde{\kappa}}. \label{d-2.1}
\end{align}
It follows that
\begin{align}
&\frac{a_{\mu}}{\kappa}=\frac{1}{M}g_{\mu 4},\; \frac{b_{\mu}}{\tilde{\kappa}}=\frac{1}{R}g_{\mu 5}, \; u^{2}=g_{44}-g_{\mu 4}g_{\mu 4}, \; v^{2}=g_{55}-g_{\mu 5}g_{\mu 5} \quad \text {and} \notag \\
 & \;uv \sin (\varphi +\psi)= g_{45}-g_{\mu 4}g_{\mu 5} \quad \text{which implies that} \quad \sin (\varphi +\psi)=\frac{1}{uv} \left( g_{45}-g_{\mu 4}g_{\mu 5}\right) .\label{d-2.2}
\end{align}

Note that,
\begin{equation}\label{2.2}
\det g=\left( g_{45}-g_{\mu 4}g_{\mu 5}\right) ^{2}-\left( g_{44}-g_{\mu 4}g_{\mu 4}\right) \left( g_{55}-g_{\nu 5}g_{\nu 5}\right)<0,
\end{equation}
i.e. using the above formulae for $g_{AB}\;$ \eqref{d-2.1} we get  $\;\det g= -u^{2}v^{2}\cos^{2} (\varphi+\psi)\;$ and $\;g_{AB}=\left( S \eta S^{T}\right)_{AB} , \; {\det S}^{2}=-\det g$ with
\begin{equation}\label{m-2}
S=\begin{bmatrix}
1 & 0 & 0 & 0 & 0 & 0 \\
0 & 1 & 0 & 0 & 0 & 0 \\
0 & 0 & 1 & 0 & 0 & 0 \\
0 & 0 & 0 & 1 & 0 & 0 \\
-\frac{M}{\kappa}a_{0} & \frac{M}{\kappa}a_{1} & \frac{M}{\kappa}a_{2} & \frac{M}{\kappa}a_{3} & \sigma & 0 \\
-\frac{R}{\tilde{\kappa}}b_{0} & \frac{R}{\tilde{\kappa}}b_{1} & \frac{R}{\tilde{\kappa}}b_{2} & \frac{R}{\tilde{\kappa}}b_{3} & \upsilon & \tau \\
\end{bmatrix},
\end{equation}
where $ \sigma = u, \;  \upsilon=v\sin(\varphi +\psi) \;$ and
$\tau= v \cos (\varphi + \psi),$ see (12) in \cite{MMM}.
\bigskip

The relations between $\tilde{M}_{AB}$ and $M_{AB}$ can be obtained from the relations for $\tilde{X}_{\mu}$ and $\hat{x}_{\mu},\;$ $\tilde{P}_{\mu}$ and $\hat{p}_{\mu},\;$ $\tilde{H}$ and $\hat{h}$ and their inverse, and can be written as  $\tilde{M}_{AB}=\left( SMS^{T}\right) _{AB}$, with the same matrix $S$ as above. We will use the metric $g_{AB}$ for lowering indices and $g^{AB}$ for raising indices, with $\tensor{g}{_A^B}_=\tensor{g}{^B_A}=\delta_{A}^{B}\;$ and $\tensor{g}{^A^B}\tensor{g}{_B_C}=\delta_{C}^{A},\;$ i.e. $\tensor{g}{^A^B}$ is the inverse of the matrix $\tensor{g}{_A_B}.$\\

\subsection{Dual $\kappa$-Minkowski spaces and $\kappa$-Poincar\'{e} algebras isomorphic to $o(1,5,g)$
}\label{sub-2}
Dual $\kappa$-Minkowski spaces and $\kappa$-Poincar\'{e} algebras can be defined by
imposing $g_{44}=0\;$ and $g_{55}=0,\;$ so that
\begin{equation}\label{2.3}
a^{2}=-u^{2}\frac{\kappa^{2}}{M^{2}}<0, \quad b^{2}=-v^{2}\frac{\tilde{\kappa}^{2}}{R^{2}}<0.
\end{equation}
Then it follows
\begin{equation}\label{2.1-d}
\left[  \tilde{X}_{\mu}, \tilde{X}_{\nu} \right]  =\frac{i}{\kappa}\left( a_{\mu}\tilde{X}_{\nu}-a_{\nu}\tilde{X}_{\mu} \right)
\end{equation}
and
\begin{equation}\label{2.2-d}
\left[  \tilde{P}_{\mu}, \tilde{P}_{\nu}\right]  =\frac{i}{\tilde{\kappa}}\left( b_{\mu}\tilde{P}_{\nu}-b_{\nu}\tilde{P}_{\mu}\right).
\end{equation}
The relation \eqref{2.1-d} defines $\kappa$-Minkowski space, while \eqref{2.2-d} describes the Born-dual $\tilde{\kappa}$-Minkowski space. Taking into account \eqref{2.3} and $g_{44}=0,\;g_{55}=0,\;g_{45}= uv \sin(\varphi+\psi)+abMR\frac{1}{\kappa\tilde{\kappa}} $, we get $ \det g=-u^{2}v^{2}\cos^{2} (\varphi +\psi)$, and we demand that $ \det g \neq 0$.
The determinant depends on the parameters $u,v$ and the angle $\varphi+\psi,\;$ but does not depend on the vectors $a_{\mu}, b_{\mu}.$ If $g_{44}=g_{55}=g_{45}=0$ and  $\det g \neq 0$, then it follows that $a^{2}\cdot b^{2}>(a\cdot b)^{2}\;$ and $a_{\mu}$ and $b_{\mu}$ cannot be collinear.
\bigskip

The algebra generated by $\tilde{X}_{\mu}, \;\tilde{P}_{\mu},\; M_{\mu\nu}$ and $\tilde{H}$ containing dual $\kappa$-Minkowski spaces and $\kappa$-Poincar\'{e} algebras  \eqref{1.1}-\eqref{1.9} is constructed from the Yang algebra \eqref{0.1}-\eqref{0.5} and is isomorphic to the $o(1,5,g)$ algebra. It depends on the parameters $M, R$, on the time-like four-vectors $\frac{a_{\mu}}{\kappa},\;\frac{b_{\mu}}{\tilde{\kappa}}\left( a^2<0, b^{2}<0\right)$ and on $g_{45}$, with the condition $\det g<0.$ In the limit when $M\rightarrow \infty$, the vector $a_{\mu}$ becomes light-like, $a^{2}=0.$
In the limit when $R\rightarrow \infty$, the vector $b_{\mu}$ becomes light-like, $b^{2}=0$ and in the limit when $M\rightarrow \infty$ and $R\rightarrow \infty$, both vectors $a_{\mu}, b_{\mu}$ become light-like. In particular, in the limit when $R\rightarrow \infty$ and $\frac{b_{\mu}}{\tilde{\kappa}}=0$ we get the $\kappa$-Minkowski space with the $\kappa$-Poincar\'{e} algebra, with $a^{2}\leq 0$ and
\begin{equation}\label{2.5}
\left[  \tilde{X}_{\mu}, \tilde{X}_{\nu} \right]  =\frac{i}{\kappa}\left( a_{\mu}\tilde{X}_{\nu}-a_{\nu}\tilde{X}_{\mu} \right) ,
\end{equation}
\begin{equation}\label{2.6}
\left[  \tilde{P}_{\mu}, \tilde{P}_{\nu}\right]=0
\end{equation}
\begin{equation}\label{2.7}
\left[  \tilde{X}_{\mu}, \tilde{P}_{\nu}\right] =i\left( \eta_{\mu\nu}\tilde{H}-\frac{1}{\kappa}a_{\nu}\tilde{P}_{\mu}\right),
\end{equation}
\begin{equation}\label{2.8}
\left[  M_{\mu\nu},\tilde{X}_{\lambda}\right]  =i\left( \eta_{\mu\lambda}\tilde{X}_{\nu}-\eta_{\nu\lambda}\tilde{X}_{\mu}+\frac{1}{\kappa}\left( a_{\mu}M_{\lambda\nu}-a_{\nu}M_{\lambda\mu}\right) \right),
\end{equation}
\begin{equation}\label{2.9}
\left[  M_{\mu\nu},\tilde{P}_{\lambda}\right]  =i\left( \eta_{\mu\lambda}\tilde{P}_{\nu}-\eta_{\nu\lambda}\tilde{P}_{\mu}\right),
\end{equation}
\begin{equation}\label{2.10}
\left[  M_{\mu\nu},\tilde{H}\right]  = \frac{-i}{\kappa}\left( a_{\nu}\tilde{P}_{\mu}-a_{\mu}\tilde{P}_{\nu}\right),
\end{equation}
\begin{equation}\label{2.11}
\left[  \tilde{H}, \tilde{X}_{\mu}\right]  =\frac{-i}{\kappa} a_{\mu}\tilde{H},
\end{equation}
\begin{equation}\label{2.12}
\left[  \tilde{H}, \tilde{P}_{\mu}\right] =0.
\end{equation}
Alternatively when $M\rightarrow \infty$ and $\frac{a_{\mu}}{\kappa}=0$,
we get the dual $\kappa$-Minkowski momentum space with $b^{2}\leq 0$.
Finally, when $M \rightarrow \infty,\; R\rightarrow \infty$ and $\frac{a_{\mu}}{\kappa}=\frac{b_{\mu}}{\tilde{\kappa}}=0$, we get the ordinary Heisenberg algebra, where $\tilde{H}$ is a central element commuting with $\tilde{X}_{\mu},\; \tilde{P}_{\mu}\;$ and $M_{\mu\nu}$, implying that $\tilde{H}$ is proportional to the identity operator $I$. The special case $M=\kappa$ and $R=\tilde{\kappa}$ is described in \cite{MMM}.

\section{Generalized Yang models isomorphic to $o(2,4,g)$}
\label {s-3}

In this section we consider the case $\epsilon_{1}=1, \; \epsilon_{2}=-1,\;$ i.e. $\eta_{AB}=diag(-1,1,1,1,1,-1)$, while the case  $\epsilon_{1}=-1, \; \epsilon_{2}=1$ is mentioned at the end of subsection \eqref{sub-3}. The Yang algebra isomorphic to $o (2,4)$ is defined as
\begin{equation}\label{3.1}
\left[  \hat{x}_{\mu},\hat{x}_{\nu}\right]  =\frac{i}{M^{2}}M_{\mu\nu}, \quad \left[  \hat{p}_{\mu},\hat{p}_{\nu}\right] =\frac{-i}{R^{2}}M_{\mu\nu},
\end{equation}
\begin{equation}\label{3.2}
\left[  M_{\mu\nu}, \hat{x}_{\lambda}\right]   =i\left( \eta_{\mu\lambda}\hat{x}_{\nu}-\eta_{\nu\lambda}\hat{x}_{\mu}\right) ,\quad  \left[  M_{\mu\nu}, \hat{p}_{\lambda}\right]   =i\left( \eta_{\mu\lambda}\hat{p}_{\nu}-\eta_{\nu\lambda}\hat{p}_{\mu}\right),
\end{equation}
\begin{equation}\label{3.3}
\left[  \hat{x}_{\mu}, \hat{p}_{\nu}\right] =i\eta_{\mu\nu}\hat{h}, \quad \left[  \hat{h}, \hat{x}_{\mu}\right]  =\frac{i}{M^{2}}\hat{p}_{\mu}, \quad \left[  \hat{h}, \hat{p}_{\mu}\right]  =\frac{i}{R^{2}}\hat{x}_{\mu},
\end{equation}
\begin{equation}\label{3.4}
\left[ M_{\mu\nu}, \hat{h} \right]  =0,
\end{equation}
\begin{equation}\label{3.5}
\left[  M_{\mu\nu},M_{\rho\sigma}\right]  =i\left( \eta_{\mu\rho}M_{\nu\sigma}-\eta_{\mu\sigma}M_{\nu\rho}-\eta_{\nu\rho}M_{\mu\sigma}+\eta_{\nu\sigma}M_{\mu\rho}\right),
\end{equation}
Let us construct new generators $\tilde{X}_{\mu},\: \tilde{P}_{\mu}$
\begin{equation}\label{3.6}
\tilde{X}_{\mu}=u\left( \cosh \varphi \hat{x}_{\mu}+\frac{R}{M}\sinh \varphi \hat{p}_{\mu}\right) +\frac{1}{\kappa} a_{\rho}M_{\mu\rho},
\end{equation}
\begin{equation}\label{3.7}
\tilde{P}_{\mu}=v\left( \cosh \psi \hat{p}_{\mu}+\frac{M}{R}\sinh \psi \hat{x}_{\mu}\right) +\frac{1}{\tilde{\kappa}} b_{\rho}M_{\mu\rho}
\end{equation}
and
\begin{equation}
\tilde{M}_{\mu \nu}=M_{\mu \nu}.
\end{equation}
The generators  $\tilde{X}_{\mu}, \;\tilde{P}_{\mu},\; M_{\mu\nu}$ and $\tilde{H}$ generate a new class of Lie algebras isomorphic to the $o(2,4)$ Yang algebra
\begin{equation}\label{3.8}
\left[  \tilde{X}_{\mu}, \tilde{X}_{\nu} \right]  =i\left( \frac{1}{M^{2}}\left(u^{2} +a^{2}\frac{M^{2}}{\kappa^{2}}\right) M_{\mu\nu}+\frac{1}{\kappa}\left( a_{\mu}\tilde{X}_{\nu}-a_{\nu}\tilde{X}_{\mu} \right)\right) ,
\end{equation}
\begin{equation}\label{3.9}
\left[  \tilde{P}_{\mu}, \tilde{P}_{\nu}\right]  =i\left( \frac{1}{R^{2}}\left(-v^{2} +b^{2}\frac{R^{2}}{\tilde{\kappa}^{2}}\right)M_{\mu\nu}+\frac{1}{\tilde{\kappa}}\left( b_{\mu}\tilde{P}_{\nu}-b_{\nu}\tilde{P}_{\mu}\right)\right) ,
\end{equation}
\begin{equation}\label{3.10}
\left[  \tilde{X}_{\mu}, \tilde{P}_{\nu}\right] =i\left( \eta_{\mu\nu}\tilde{H}+\frac{1}{\tilde{\kappa}} b_{\mu}\tilde{X}_{\nu}-\frac{1}{\kappa}a_{\nu}\tilde{P}_{\mu}+\frac{1}{MR} \left( uv \sinh(\psi-\varphi)+abMR\frac{1}{\kappa\tilde{\kappa}}\right) M_{\mu\nu}\right) ,
\end{equation}
\begin{equation}\label{3.11}
\left[  M_{\mu\nu},\tilde{X}_{\lambda}\right]  =i\left( \eta_{\mu\lambda}\tilde{X}_{\nu}-\eta_{\nu\lambda}\tilde{X}_{\mu}+\frac{1}{\kappa}\left( a_{\mu}M_{\lambda\nu}-a_{\nu}M_{\lambda\mu}\right) \right),
\end{equation}
\begin{equation}\label{3.12}
\left[  M_{\mu\nu},\tilde{P}_{\lambda}\right]  =i\left( \eta_{\mu\lambda}\tilde{P}_{\nu}-\eta_{\nu\lambda}\tilde{P}_{\mu}+\frac{1}{\tilde{\kappa}}\left( b_{\mu}M_{\lambda\nu}-b_{\nu}M_{\lambda\mu}\right) \right),
\end{equation}
where
\begin{equation}\label{3.13}
\tilde{H}=\hat{h}uv\cosh(\psi-\varphi)+\frac{1}{\kappa} a\tilde{P}-\frac{1}{\tilde{\kappa}} b \tilde{X}-\frac{1}{\kappa\tilde{\kappa}} a_{\rho}b_{\sigma}M_{\rho\sigma}
\end{equation}
and
\begin{equation}\label{3.14}
\left[  M_{\mu\nu},\tilde{H}\right]  =i\left( \frac{1}{\tilde{\kappa}}\left( b_{\nu}\tilde{X}_{\mu}-b_{\mu}\tilde{X}_{\nu}\right) -\frac{1}{\kappa}\left( a_{\nu}\tilde{P}_{\mu}-a_{\mu}\tilde{P}_{\nu}\right)\right) ,
\end{equation}
\begin{equation}\label{3.15}
\left[  \tilde{H}, \tilde{X}_{\mu}\right]  =i\left(\frac{1}{M^{2}}\left(u^{2} +a^{2}\frac{M^{2}}{\kappa^{2}}\right) \tilde{P}_{\mu}-\frac{1}{MR} \left( uv \sinh(\psi-\varphi)+abMR\frac{1}{\kappa\tilde{\kappa}}\right)\tilde{X}_{\mu}-\frac{1}{\kappa} a_{\mu}\tilde{H}\right),
\end{equation}
\begin{equation}\label{3.16}
\left[  \tilde{H}, \tilde{P}_{\mu}\right] =-i\left(\frac{1}{R^{2}}\left(-v^{2} +b^{2}\frac{R^{2}}{\tilde{\kappa}^{2}}\right) \tilde{X}_{\mu}-\frac{1}{MR} \left( uv \sinh(\psi-\varphi)+abMR\frac{1}{\kappa\tilde{\kappa}}\right)\tilde{P}_{\mu}+\frac{1}{\tilde{\kappa}} b_{\mu}\tilde{H}\right).
\end{equation}
The Born duality holds for
$M \leftrightarrow R, \;\kappa \leftrightarrow \tilde{\kappa},\; a_{\mu}\rightarrow -b_{\mu},\; b_{\mu}\rightarrow a_{\mu},\; u \leftrightarrow v,\; \varphi \rightarrow -\psi,\; \psi \rightarrow -\varphi,\; \tilde{X}_{\mu} \rightarrow  -\tilde{P}_{\mu},\; \tilde{P}_{\mu} \rightarrow \tilde{X}_{\mu},\; M_{\mu\nu}\leftrightarrow M_{\mu\nu},\; \tilde{H} \leftrightarrow \tilde{H}$ and $ \epsilon_{1}\leftrightarrow \epsilon_{2}$.
If $\frac{a_{\mu}}{\kappa}=0\;$ and $\frac{b_{\mu}}{\tilde{\kappa}}=0$ then the above algebra reduces to a Khruschev-Leznov model \cite{KL} isomorphic to the $o(2,4)$ algebra, with  $\rho=uv \sinh (\psi-\varphi).$\\
If we define
\begin{equation}
\tilde{M}_{\mu 4}=M \tilde{X}_{\mu}, \; \tilde{M}_{\mu 5}=R\tilde{P}_{\mu}, \;  \tilde{M}_{45}=MR \tilde{H}
\end{equation}
as in \eqref{2-d} then the algebra \eqref{3.8}-\eqref{3.16} becomes an $o(2,4,g)$ algebra and  $\left[  \tilde{M}_{A B},\tilde{M}_{C D}\right]$ is given in \eqref{2.1}, with the metric $g_{AB}$ given by a symmetric matrix defined as
\begin{align}\label{3-d}
 &g_{\mu\nu}=\eta_{\mu\nu}, \; g_{\mu 4}=\frac{M}{\kappa}a_{\mu}, \; g_{\mu 5}=\frac{R}{\tilde{\kappa}}b_{\mu}, \; g_{44}=u^{2} +a^{2}\frac{M^{2}}{\kappa^{2}}, \; g_{55}=-v^{2} +b^{2}\frac{R^{2}}{\tilde{\kappa}^{2}},\; \text{ and } \notag\\  &g_{45}=g_{54}= uv \sinh(\psi-\varphi)+abMR\frac{1}{\kappa\tilde{\kappa}}, \; A, B, C, D=0,1,2,3,4,5.
 \end{align}
The inverse relations are given by
\begin{equation}\label{d-3.1}
u^{2}=g_{44}-g_{\mu 4}g_{\mu 4}, \;v^{2}=g_{\mu 5}g_{\mu 5}-g_{55} \quad \text{and} \quad \sinh (\psi-\varphi)=\frac{1}{uv}\left( g_{45}-g_{\mu 4}g_{\mu 5}\right).
\end{equation}
Note that,
\begin{equation}\label{3.17}
\det g=\left( g_{45}-g_{\mu 4}g_{\mu 5}\right) ^{2}-\left( g_{44}-g_{\mu 4}g_{\mu 4}\right) \left( g_{55}-g_{\nu 5}g_{\nu 5}\right) \geq 1,
\end{equation}
 and $\;g_{AB}=\left( S \eta S^{T}\right)_{AB}, \; {\det S}^{2}=\det g$ with
\begin{equation}
S=\begin{bmatrix}
1 & 0 & 0 & 0 & 0 & 0 \\
0 & 1 & 0 & 0 & 0 & 0 \\
0 & 0 & 1 & 0 & 0 & 0 \\
0 & 0 & 0 & 1 & 0 & 0 \\
-\frac{M}{\kappa}a_{0} & \frac{M}{\kappa}a_{1} & \frac{M}{\kappa}a_{2} & \frac{M}{\kappa}a_{3} & \sigma & 0 \\
-\frac{R}{\tilde{\kappa}}b_{0} & \frac{R}{\tilde{\kappa}}b_{1} & \frac{R}{\tilde{\kappa}}b_{2} & \frac{R}{\tilde{\kappa}}b_{3} & \upsilon & \tau \\
\end{bmatrix},
\end{equation}
where $ \sigma = u, \;  \upsilon=v\sinh(\psi-\varphi) \;$ and
$\tau= v \cosh (\psi-\varphi ).$

The relations between $\tilde{M}_{AB}$ and $M_{AB}$ can be written as  $\tilde{M}_{AB}=\left( SMS^{T}\right) _{AB}$ with the matrix $S$ as above.
\subsection{Dual $\kappa$-Minkowski spaces and $\kappa$-Poincar\'{e} algebras isomorphic to $o(2,4,g)$
}\label{sub-3}
If we impose $g_{44}=0\;$ and $g_{55}=0,\;$ we obtain
\begin{equation}\label{3.18}
a^{2}=-u^{2}\frac{\kappa^{2}}{M^{2}}<0, \quad b^{2}=v^{2}\frac{\tilde{\kappa}^{2}}{R^{2}}>0
\end{equation}
and $\left[  \tilde{X}_{\mu}, \tilde{X}_{\nu} \right], \; \left[  \tilde{P}_{\mu}, \tilde{P}_{\nu} \right]$ are as in subsection \eqref{sub-2}.
Taking into account \eqref{3.18} and $g_{44}=0,\;g_{55}=0,\;g_{45}= uv \sinh(\psi-\varphi)+abMR\frac{1}{\kappa\tilde{\kappa}} $, we get $ \det g=u^{2}v^{2}\cosh^{2} (\psi-\varphi) \geq 1$,
which depends on the parameters $u,v$ and  $\psi-\varphi,\;$ but does not depend on the vectors $a_{\mu}, b_{\mu}.$\\
\bigskip

Alternatively if $\epsilon_{1}=-1,\; \epsilon_{2}=1$ we get a new class of algebras isomorphic to $o(2,4,g)$ with a metric
 \begin{align}
 &g_{\mu\nu}=\eta_{\mu\nu}, \; g_{\mu 4}=\frac{M}{\kappa}a_{\mu}, \; g_{\mu 5}=\frac{R}{\tilde{\kappa}}b_{\mu}, \; g_{44}=-u^{2} +a^{2}\frac{M^{2}}{\kappa^{2}}, \; g_{55}=v^{2} +b^{2}\frac{R^{2}}{\tilde{\kappa}^{2}},\; \text{ and} \notag\\  &g_{45}=g_{54}=- uv \sinh(\psi-\varphi)+ab\frac{MR}{\kappa\tilde{\kappa}}.
 \end{align}
Now if we impose $g_{44}=0\;$ and $g_{55}=0,\;$ we obtain
\begin{equation}\label{3.19}
a^{2}=u^{2}\frac{\kappa^{2}}{M^{2}}>0, \quad b^{2}=-v^{2}\frac{\tilde{\kappa}^{2}}{R^{2}}<0.
\end{equation}
Taking into  account \eqref{3.19} and $g_{44}=0,\;g_{55}=0,\;g_{45}= -uv \sinh(\psi-\varphi)+abMR\frac{1}{\kappa\tilde{\kappa}} $ we get\ $ \det g=u^{2}v^{2}\cosh^{2} (\psi-\varphi)\geq 1$.

\section{Generalized Yang models isomorphic to $o(3,3,g)$}
\label {s-4}

In this section  we consider the case $\epsilon_{1}=\epsilon_{2}=-1,\;$ i.e. $\eta_{AB}=(-1,1,1,1,-1,-1)$.
The Yang algebra isomorphic to $o(3,3)$ is defined as
\begin{equation}\label{4.1}
\left[  \hat{x}_{\mu},\hat{x}_{\nu}\right]  =\frac{-i}{M^{2}}M_{\mu\nu}, \quad \left[  \hat{p}_{\mu},\hat{p}_{\nu}\right] =\frac{-i}{R^{2}}M_{\mu\nu},
\end{equation}
\begin{equation}\label{4.2}
\left[  M_{\mu\nu}, \hat{x}_{\lambda}\right]   =i\left( \eta_{\mu\lambda}\hat{x}_{\nu}-\eta_{\nu\lambda}\hat{x}_{\mu}\right) ,\quad  \left[  M_{\mu\nu}, \hat{p}_{\lambda}\right]   =i\left( \eta_{\mu\lambda}\hat{p}_{\nu}-\eta_{\nu\lambda}\hat{p}_{\mu}\right),
\end{equation}
\begin{equation}\label{4.3}
\left[  \hat{x}_{\mu}, \hat{p}_{\nu}\right] =i\eta_{\mu\nu}\hat{h}, \quad \left[  \hat{h}, \hat{x}_{\mu}\right]  =\frac{-i}{M^{2}}\hat{p}_{\mu}, \quad \left[  \hat{h}, \hat{p}_{\mu}\right]  =\frac{i}{R^{2}}\hat{x}_{\mu},
\end{equation}
\begin{equation}\label{4.4}
\left[ M_{\mu\nu}, \hat{h} \right]  =0,
\end{equation}
\begin{equation}\label{4.5}
\left[  M_{\mu\nu},M_{\rho\sigma}\right]  =i\left( \eta_{\mu\rho}M_{\nu\sigma}-\eta_{\mu\sigma}M_{\nu\rho}-\eta_{\nu\rho}M_{\mu\sigma}+\eta_{\nu\sigma}M_{\mu\rho}\right),
\end{equation}
Let us construct the new generators
\begin{equation}\label{4.6}
\tilde{X}_{\mu}=u\left( \cos \varphi \hat{x}_{\mu}+\frac{R}{M}\sin \varphi \hat{p}_{\mu}\right) +\frac{1}{\kappa} a_{\rho}M_{\mu\rho},
\end{equation}
\begin{equation}\label{4.7}
\tilde{P}_{\mu}=v\left( \cos \psi \hat{p}_{\mu}+\frac{M}{R}\sin \psi \hat{x}_{\mu}\right) +\frac{1}{\tilde{\kappa}} b_{\rho}M_{\mu\rho}.
\end{equation}
The generators $\tilde{X}_{\mu}, \;\tilde{P}_{\mu},\; M_{\mu\nu}$ and $\tilde{H}$ generate a new class of Lie algebras isomorphic to the $o(3,3)$ algebra
\begin{equation}\label{4.8}
\left[  \tilde{X}_{\mu}, \tilde{X}_{\nu} \right]  =i\left( \frac{1}{M^{2}}\left(-u^{2} +a^{2}\frac{M^{2}}{\kappa^{2}}\right) M_{\mu\nu}+\frac{1}{\kappa}\left( a_{\mu}\tilde{X}_{\nu}-a_{\nu}\tilde{X}_{\mu} \right)\right) ,
\end{equation}
\begin{equation}\label{4.9}
\left[  \tilde{P}_{\mu}, \tilde{P}_{\nu}\right]  =i\left( \frac{1}{R^{2}}\left(-v^{2} +b^{2}\frac{R^{2}}{\tilde{\kappa}^{2}}\right)M_{\mu\nu}+\frac{1}{\tilde{\kappa}}\left( b_{\mu}\tilde{P}_{\nu}-b_{\nu}\tilde{P}_{\mu}\right)\right) ,
\end{equation}
\begin{equation}\label{4.10}
\left[  \tilde{X}_{\mu}, \tilde{P}_{\nu}\right] =i\left( \eta_{\mu\nu}\tilde{H}+\frac{1}{\tilde{\kappa}} b_{\mu}\tilde{X}_{\nu}-\frac{1}{\kappa}a_{\nu}\tilde{P}_{\mu}+\frac{1}{MR} \left( -uv \sin(\varphi+\psi)+abMR\frac{1}{\kappa\tilde{\kappa}}\right) M_{\mu\nu}\right) ,
\end{equation}
\begin{equation}\label{4.11}
\left[  M_{\mu\nu},\tilde{X}_{\lambda}\right]  =i\left( \eta_{\mu\lambda}\tilde{X}_{\nu}-\eta_{\nu\lambda}\tilde{X}_{\mu}+\frac{1}{\kappa}\left( a_{\mu}M_{\lambda\nu}-a_{\nu}M_{\lambda\mu}\right) \right),
\end{equation}
\begin{equation}\label{4.12}
\left[  M_{\mu\nu},\tilde{P}_{\lambda}\right]  =i\left( \eta_{\mu\lambda}\tilde{P}_{\nu}-\eta_{\nu\lambda}\tilde{P}_{\mu}+\frac{1}{\tilde{\kappa}}\left( b_{\mu}M_{\lambda\nu}-b_{\nu}M_{\lambda\mu}\right) \right),
\end{equation}
where
\begin{equation}\label{4.13}
\tilde{H}=\hat{h}uv\cos(\varphi+\psi)+\frac{1}{\kappa} a\tilde{P}-\frac{1}{\tilde{\kappa}} b \tilde{X}-\frac{1}{\kappa\tilde{\kappa}} a_{\rho}b_{\sigma}M_{\rho\sigma}
\end{equation}
and
\begin{equation}\label{4.14}
\left[  M_{\mu\nu},\tilde{H}\right]  =i\left( \frac{1}{\tilde{\kappa}}\left( b_{\nu}\tilde{X}_{\mu}-b_{\mu}\tilde{X}_{\nu}\right) -\frac{1}{\kappa}\left( a_{\nu}\tilde{P}_{\mu}-a_{\mu}\tilde{P}_{\nu}\right)\right) ,
\end{equation}
\begin{equation}\label{4.15}
\left[  \tilde{H}, \tilde{X}_{\mu}\right]  =i\left(\frac{1}{M^{2}}\left(-u^{2} +a^{2}\frac{M^{2}}{\kappa^{2}}\right) \tilde{P}_{\mu}-\frac{1}{MR} \left( -uv \sin(\varphi+\psi)+abMR\frac{1}{\kappa\tilde{\kappa}}\right)\tilde{X}_{\mu}-\frac{1}{\kappa} a_{\mu}\tilde{H}\right),
\end{equation}
\begin{equation}\label{4.16}
\left[  \tilde{H}, \tilde{P}_{\mu}\right] =-i\left(\frac{1}{R^{2}}\left(-v^{2} +b^{2}\frac{R^{2}}{\tilde{\kappa}^{2}}\right) \tilde{X}_{\mu}-\frac{1}{MR} \left( -uv \sin(\varphi+\psi)+abMR\frac{1}{\kappa\tilde{\kappa}}\right)\tilde{P}_{\mu}+\frac{1}{\tilde{\kappa}} b_{\mu}\tilde{H}\right).
\end{equation}
The Born duality holds as in sections \ref{s-2} and \ref{s-3}. If $\frac{a_{\mu}}{\kappa}=0\;$ and $\frac{b_{\mu}}{\tilde{\kappa}}=0$ then the algebra \eqref{4.8}-\eqref{4.16} reduces to a model of Khruschev-Leznov type \cite {KL} isomorphic to an $o(3,3)$ algebra, with $\rho=-uv \sin (\varphi+\psi).$\\
If we define
\begin{equation}\label{4.17}
\tilde{M}_{\mu 4}=M \tilde{X}_{\mu}, \; \tilde{M}_{\mu 5}=R\tilde{P}_{\mu}, \;  \tilde{M}_{45}=MR \tilde{H}
\end{equation}
then the algebra \eqref{4.8}-\eqref{4.16} becomes the $o(3,3,g)$ algebra and $\left[  \tilde{M}_{A B},\tilde{M}_{C D}\right]$ is as in sections \ref{s-2} and \ref{s-3}, with the metric  $g_{AB}$ given by
\begin{align}
&g_{\mu\nu}=\eta_{\mu\nu}, \; g_{\mu 4}=\frac{M}{\kappa}a_{\mu}, \; g_{\mu 5}=\frac{R}{\tilde{\kappa}}b_{\mu}, \; g_{44}=-u^{2} +a^{2}\frac{M^{2}}{\kappa^{2}}, \; g_{55}=-v^{2} +b^{2}\frac{R^{2}}{\tilde{\kappa}^{2}},\;\notag  \\
& g_{45}=g_{54}= -uv \sin(\varphi+\psi)+abMR\frac{1}{\kappa\tilde{\kappa}}, \; A, B, C, D=0,1,2,3,4,5. \label{4.18}
\end{align}
Note that $\det g<0$ and ${\det S}^{2}=-\det g$ with $S$ given in \eqref{m-2} where $ \sigma = u, \;  \upsilon=v\sin(\varphi +\psi) \;$ and $\tau= v \cos (\varphi + \psi).$.
\subsection{Dual $\kappa$-Minkowski spaces and $\kappa$-Poincar\'{e} algebras isomorphic to $o(3,3,g)$
}\label{sub-4}
When we impose $g_{44}=0\;$ and $g_{55}=0,\;$ we obtain
\begin{equation}\label{4.19}
a^{2}=u^{2}\frac{\kappa^{2}}{M^{2}}>0, \quad b^{2}=v^{2}\frac{\tilde{\kappa}^{2}}{R^{2}}>0
\end{equation}
 and we get $\left[  \tilde{X}_{\mu}, \tilde{X}_{\nu} \right],\;
\left[  \tilde{P}_{\mu}, \tilde{P}_{\nu}\right] $ as in subsections \eqref{sub-2} and \eqref{sub-3}: then $\left[  \tilde{X}_{\mu}, \tilde{X}_{\nu} \right]$ defines the $\kappa$-Minkowski space and $\left[  \tilde{P}_{\mu}, \tilde{P}_{\nu}\right]$  the Born dual $\tilde{\kappa}$-Minkowski space. In particular, when  $g_{44}=0,\;g_{55}=0,\;g_{45}= -uv \sin(\varphi+\psi)+abMR\frac{1}{\kappa\tilde{\kappa}} $ we get $ \det g=-u^{2}v^{2}\cos^{2} (\varphi +\psi)$ and  $ \det g \neq 0$.
The determinant depends on the parameters $u,v$ and $\varphi + \psi$ but not on the vectors $a_{\mu}$ and $b_{\mu}$.
If $g_{44}=0,\;g_{55}=0,\;g_{45}=0,\; \det g<0$ we get $\left( ab\right) ^{2}-a^{2}b^{2}<0\;$, i.e. $\;a^{2}b^{2}>0$ and the vectors $a_{\mu}$ and $b_{\mu}$ cannot be collinear.\\
The algebra generated by $\tilde{X}_{\mu}, \;\tilde{P}_{\mu},\; \tilde{H}$ and $M_{\mu\nu}$ containing dual $\kappa$-Minkowski spaces is constructed from the Yang algebra and  is isomorphic to the $o(3,3,g)$ algebra. It depends on the parameters $M,R$ and space-like vectors $\frac{a_{\mu}}{\kappa}, \; \frac{b_{\mu}}{\tilde{\kappa}},\; a^{2}>0,\; b^{2}>0$, and $g_{45}$, with $\det g<0$.
\bigskip

\subsection{Unification of the four cases $\epsilon_{1} = \pm 1,\; \epsilon_{2} = \pm 1$}\label{sub-4.2}

All four cases $\epsilon_{1}=\pm 1,\; \epsilon_{2}=\pm 1$ considered in sections \ref{s-2}, \ref{s-3} and \ref{s-4} can be unified in the following way:
\begin{align}
&g_{\mu\nu}=\eta_{\mu\nu}, \; g_{\mu 4}=\frac{M}{\kappa}a_{\mu}, \; g_{\mu 5}=\frac{R}{\tilde{\kappa}}b_{\mu}, \notag\\
& \; g_{44}=\epsilon_{1}u^{2} +a^{2}\frac{M^{2}}{\kappa^{2}}, \; g_{55}=\epsilon_{2}v^{2} +b^{2}\frac{R^{2}}{\tilde{\kappa}^{2}},\;\notag  \\
& g_{45}=g_{54}= \sqrt{\epsilon_{1}\epsilon_{2}}uv \sin\left( \sqrt{\epsilon_{1}\epsilon_{2}}(\epsilon_{1}\varphi+\epsilon_{2}\psi)\right) +abMR\frac{1}{\kappa\tilde{\kappa}},\notag\\
& \det g=-\epsilon_{1}\epsilon_{2}\cos^{2}\left( \sqrt{\epsilon_{1}\epsilon_{2}}(\epsilon_{1}\varphi+\epsilon_{2}\psi)\right). \label{4.20}
\end{align}
The cases $g_{\mu 4}=g_{\mu 5}=0, \; g_{ 45} \neq 0, \; \epsilon_{1}=\pm 1,\; \epsilon_{2}=\pm 1$ correspond to the Khruschev-Leznov  type of models given in \cite{KL}.\\
The corresponding algebras  containing dual $\kappa$-Minkowski spaces and $\kappa$-Poincar\'{e} algebras are defined as
\begin{equation}\label{4.21}
\left[  \tilde{X}_{\mu}, \tilde{X}_{\nu} \right]  =\frac{i}{\kappa}\left( a_{\mu}\tilde{X}_{\nu}-a_{\nu}\tilde{X}_{\mu} \right) ,
\end{equation}
\begin{equation}\label{4.22}
\left[  \tilde{P}_{\mu}, \tilde{P}_{\nu}\right]  =\frac{i}{\tilde{\kappa}}\left( b_{\mu}\tilde{P}_{\nu}-b_{\nu}\tilde{P}_{\mu}\right) ,
\end{equation}
\begin{equation}\label{4.23}
\left[  \tilde{X}_{\mu}, \tilde{P}_{\nu}\right] =i\left( \eta_{\mu\nu}\tilde{H}+\frac{1}{\tilde{\kappa}} b_{\mu}\tilde{X}_{\nu}-\frac{1}{\kappa}a_{\nu}\tilde{P}_{\mu}+\left( \frac{\sqrt{\epsilon_{1}\epsilon_{2}}uv}{MR}  \sin\left( \sqrt{\epsilon_{1}\epsilon_{2}}(\epsilon_{1}\varphi+\epsilon_{2}\psi)\right) +\frac{ab}{\kappa\tilde{\kappa}}\right) M_{\mu\nu}\right) ,
\end{equation}
\begin{equation}\label{4.24}
\left[  M_{\mu\nu},\tilde{X}_{\lambda}\right]  =i\left( \eta_{\mu\lambda}\tilde{X}_{\nu}-\eta_{\nu\lambda}\tilde{X}_{\mu}+\frac{1}{\kappa}\left( a_{\mu}M_{\lambda\nu}-a_{\nu}M_{\lambda\mu}\right) \right),
\end{equation}
\begin{equation}\label{4.25}
\left[  M_{\mu\nu},\tilde{P}_{\lambda}\right]  =i\left( \eta_{\mu\lambda}\tilde{P}_{\nu}-\eta_{\nu\lambda}\tilde{P}_{\mu}+\frac{1}{\tilde{\kappa}}\left( b_{\mu}M_{\lambda\nu}-b_{\nu}M_{\lambda\mu}\right) \right),
\end{equation}
\begin{equation}\label{4.26}
\left[  M_{\mu\nu},\tilde{H}\right]  =i\left( \frac{1}{\tilde{\kappa}}\left( b_{\nu}\tilde{X}_{\mu}-b_{\mu}\tilde{X}_{\nu}\right) -\frac{1}{\kappa}\left( a_{\nu}\tilde{P}_{\mu}-a_{\mu}\tilde{P}_{\nu}\right)\right) ,
\end{equation}
\begin{equation}\label{4.27}
\left[  \tilde{H}, \tilde{X}_{\mu}\right]  =-i\left( \left(\frac{\sqrt{\epsilon_{1}\epsilon_{2}}}{MR} uv \sin\left( \sqrt{\epsilon_{1}\epsilon_{2}}(\epsilon_{1}\varphi+\epsilon_{2}\psi)\right) +\frac{ab}{\kappa\tilde{\kappa}}\right)\tilde{X}_{\mu}+\frac{1}{\kappa} a_{\mu}\tilde{H}\right),
\end{equation}
\begin{equation}\label{4.28}
\left[  \tilde{H}, \tilde{P}_{\mu}\right] =i\left( \left(\frac{\sqrt{\epsilon_{1}\epsilon_{2}}}{MR} uv \sin\left( \sqrt{\epsilon_{1}\epsilon_{2}}(\epsilon_{1}\varphi+\epsilon_{2}\psi)\right) +\frac{ab}{\kappa\tilde{\kappa}}\right)\tilde{P}_{\mu}-\frac{1}{\tilde{\kappa}} b_{\mu}\tilde{H}\right),
\end{equation}
where
\begin{equation}\label{4.29}
\tilde{H}=\hat{h}\det S+\frac{1}{\kappa} a\tilde{P}-\frac{1}{\tilde{\kappa}} b \tilde{X}-\frac{1}{\kappa\tilde{\kappa}} a_{\rho}b_{\sigma}M_{\rho\sigma}
\end{equation}
and
\begin{equation}\label{4.30}
a^{2}=-\epsilon_{1}u^{2}\frac{\kappa^{2}}{M^{2}}, \quad b^{2}=-\epsilon_{2}v^{2}\frac{\tilde{\kappa}^{2}}{R^{2}}.
\end{equation}
Generally if $\varphi=\psi=0$ then the algebra \eqref{4.8}-\eqref{4.16} depends only on the parameters $\frac{a_{\mu}}{\kappa}\;$ and $\;\frac{b_{\mu}}{\tilde{\kappa}}$.
The case $a^{2}<0$, $b^{2}<0$ corresponds to $\epsilon_{1}=\epsilon_{2}=1$  and is related to the $o(1,5) $ algebra. The cases $a^{2}<0, \; b^{2}>0$ and $a^{2}>0, \; b^{2}<0$ correspond to $\epsilon_{1}=1,\; \epsilon_{2}=-1$ and $\epsilon_{1}=-1,\; \epsilon_{2}=1$, respectively, and are related to the $o(2,4)$ algebra. The case $a^{2}>0, \; b^{2}>0$, that corresponds to  $\epsilon_{1}=\epsilon_{2}=-1$, is related to the $o(3,3)$ algebra. The case $a^{2}=0$ implies $\;M \rightarrow \infty \;$ i.e. $\;\left[ \hat{x}_{\mu}, \hat{x}_{\nu}\right] =0\;$ and the case $b^{2}=0$ implies $\;R \rightarrow \infty \;$ i.e. $\;\left[ \hat{p}_{\mu}, \hat{p}_{\nu}\right] =0\;$.\\
\bigskip

In particular, if $\tilde{X}_{\mu}= \hat{x}_{\mu} + \frac{1}{\kappa}a_{\rho}M_{\mu\rho}, \; \tilde{P}_{\mu}=\hat{p}_{\mu},\;
\tilde{H}=\hat{h} +\frac{1}{\kappa}a\hat{p}\;$ i.e. $u=v=1,\; \varphi=\psi=0$ and $\frac{b_{\mu}}{\tilde{\kappa}}=0$, the above algebra becomes the algebra \eqref{2.5}-\eqref{2.12}. This algebra coincides with the construction of the natural realizations for the $\kappa$-Poincar\'{e} algebra presented in \cite{MK}. In addition, if $\frac{a_{\mu}}{\kappa}=0$ we get the ordinary Heisenberg algebra with the Lorentz algebra  where $\tilde{H}$ is a central element commuting with $\tilde{X}_{\mu},\; \tilde{P}_{\mu}\;$ and $\;M_{\mu\nu}\;$ and  is proportional to the identity operator $I$.

\section{Weyl realization for generalized Yang model and dual $\kappa$-Minkowski spaces}\label{s-5}
In this section we consider Weyl realization for orthogonal algebras $o(1,5),\; o(2,4),$ $o(3,3)$ and $o(1,5,g),\; o(2,4,g),\; o(3,3,g)$ and for dual $\kappa$-Minkowski spaces, using a formalism analogue to that of \cite{MM-4} \cite{Lukierski-3}.

\subsection{Weyl realizations for $o(1,5),\; o(2,4), \; o(3,3)$ algebras}
\label{sub-5.1}
Let us consider the algebra defined as
\begin{equation}
\left[  M_{A B},M_{C D}\right]  =i\tensor{C}{_A_B_,_C_D^E^F}M_{EF}=i\left( \eta_{A C}M_{B D}-\eta_{A D}M_{B C}+\eta_{B D}M_{A C}-\eta_{B C}M_{A D}\right),
\end{equation}
where $ \eta_{AB}=\text{diag}(-1,1,1,1,\epsilon_{1},\epsilon_{2})$,
corresponding to the algebra $o(1,5)$ for $\epsilon_{1}=\epsilon_{2}=1,\;$ to $o(2,4)$ for $\epsilon_{1}=1,\; \epsilon_{2}=-1$ or $\epsilon_{1}=-1, \;\epsilon_{2}=1$, and to $o(3,3)$ for $\epsilon_{1}=\epsilon_{2}=-1$.\\
\bigskip

Starting with the generalized Heisenberg algebra defined with the commutative coordinates $x_{AB}$ and their corresponding momenta $k^{AB}$
\begin{equation}
\left[ x_{AB}, x_{CD}\right] =\left[ k^{AB}, k^{CD}\right] =0, \quad \left[ x_{AB}, k^{CD}\right] =i\left( \tensor{\delta}{_A^C}\tensor{\delta}{_B^D}-\tensor{\delta}{_A^D}\tensor{\delta}{_B^C}\right),
\end{equation}
where $k^{CD}=\eta^{CM}\eta^{DN}k_{MN}$ and using the general Weyl realization of a Lie algebras corresponding to the Weyl symmetric ordering \cite{Meljanac-3} \cite{Meljanac-4} we express the generators $M_{AB}$ as
\begin{equation}
M_{AB}=x_{CD}\tensor{\left( \frac{\mathbf{C}}{1-e^{-\mathbf{C}}}\right) }{_A_B^C^D},
\end{equation}
where $\tensor{\mathbf{C}}{_A_B^C^D}=-\frac{1}{2}\tensor{C}{_A_B_,_E_F^C^D}k^{EF}$ and
\begin{equation}
\tensor{C}{_A_B_,_E_F^C^D}=\frac{1}{2}\left[ -\eta_{BE}\left( \tensor{\delta}{_A^C}\tensor{\delta}{_F^B}-\tensor{\delta}{_A^D}\tensor{\delta}{_F^C}\right) +\eta_{AF}\left( \tensor{\delta}{_E^C}\tensor{\delta}{_B^D}-\tensor{\delta}{_E^D}\tensor{\delta}{_B^C}\right) -(A \leftrightarrow B)\right].
\end{equation}
The structure constants $\tensor{C}{_A_B_,_E_F^C^D}$ are multiplied with $\hbar$, in our convention $\hbar=1$ and in the classical limit when $\hbar=0$ all the generators commute. The Weyl realization of $M_{AB}$ in terms of the generalized Heisenberg algebra generated with $x_{AB}$ and $k^{AB}$ reads up to second order,
\begin{equation}\label{5.5}
M_{AB}=x_{AB}+\frac{1}{2}x_{CD}\tensor{\mathbf{C}}{_A_B^C^D}+\frac{1}{12}x_{CD}\tensor{\left( \mathbf{C}^{2}\right) }{_A_B^C^D}
\end{equation}
where
\begin{equation}\label{5.6}
\tensor{\mathbf{C}}{_A_B^E^F}=\frac{1}{2}\left( \tensor{\delta}{_A^E}\tensor{k}{_B^F}+\tensor{\delta}{_B^F}\tensor{k}{_A^E}-(E\leftrightarrow F)\right),
\end{equation}
\begin{equation}\label{5.7}
\tensor{\left( \mathbf{C}^{2}\right) }{_A_B^E^F}=\frac{1}{2}\left( 2\tensor{k}{_A^E}\tensor{k}{_B^F}+\tensor{\delta}{_B^F}k_{AC}k^{CE}+\tensor{\delta}{_A^E}k_{BC}k^{CF}-(E\leftrightarrow F)\right),
\end{equation}
and the indices are lowered and raised by the means of the metric $\eta_{AB}$ and $\eta^{AB}$ respectively. Inserting $\mathbf{C}$ in \eqref{5.5} we find up to first order
\begin{equation}
M_{AB}=x_{AB}+\frac{1}{2}\left( x_{AE}\tensor{k}{_B^E}- x_{BE}\tensor{k}{_A^E}\right)
\end{equation}
and
\begin{equation}
\left[ M_{AB}, k^{CD}\right] =i\left( \tensor{\delta}{_A^C}\tensor{\delta}{_B^D}-\tensor{\delta}{_A^D}\tensor{\delta}{_B^C}\right)+\frac{i}{2}\left( \tensor{\delta}{_A^C}\tensor{k}{_B^D}-\tensor{\delta}{_B^C}\tensor{k}{_A^D}+\tensor{\delta}{_B^D}\tensor{k}{_A^C}-\tensor{\delta}{_A^D}\tensor{k}{_B^C}\right).
\end{equation}
We can write $x_{AB}$ and $k^{AB}$ in the terms of the four dimensional variables
\begin{align}
& k^{\mu 4}=\frac{q^{\mu}}{M},\quad k^{\mu 5}=\frac{y^{\mu}}{R}, \quad k^{45}=\frac{w}{MR},\\
&x_{\mu 4}=M x_{\mu}, \quad x_{\mu 5}=R p_{\mu}, \quad x_{45}=MRh,
\end{align}
so that
\begin{equation}
\left[ x_{\mu},q_{\nu}\right] =i\delta_{\mu}^{\nu}, \quad \left[ p_{\mu},y_{\nu}\right] =i\delta_{\mu}^{\nu}, \quad \left[ h,w\right] =i.
\end{equation}
\bigskip

\subsection{Weyl realizations for $o(1,5, g),\; o(2,4, g)\;$  and  $\; o(3,3, g)\;$ algebras}
\label{sub-5.2}
\bigskip
The algebras $o(1,5, g),\; o(2,4,g)\;$  and  $\; o(3,3, g)\;$ generated by the $\tilde{M}_{AB}$ are defined as
\begin{equation}
\left[ \tilde{ M}_{A B},\tilde{M}_{C D}\right]  =i\tensor{\tilde{C}}{_A_B_,_C_D^E^F}\tilde{M}_{EF}=i\left( g_{A C}\tilde{M}_{B D}-g_{A D}\tilde{M}_{B C}+g_{B D}\tilde{M}_{A C}-g_{B C}\tilde{M}_{A D}\right),
\end{equation}
with $g_{AB}=\left( S \eta S^{T}\right) _{AB}$, corresponding to  $o(1,5, g),\; o(2,4, g),\; o(3,3, g).$
As in the previous subsection, starting  with the generalized Heisenberg algebra defined with the commutative coordinates $X_{AB}$ and their corresponding momenta $K^{AB}$
\begin{equation}
\left[ X_{AB}, X_{CD}\right] =\left[ K^{AB}, K^{CD}\right] =0, \quad \left[ X_{AB}, K^{CD}\right] =i\left( \tensor{\delta}{_A^C}\tensor{\delta}{_B^D}-\tensor{\delta}{_A^D}\tensor{\delta}{_B^C}\right),
\end{equation}
where $K^{CD}=g^{CM}g^{DN}K_{MN}$, and using the general Weyl realization of the Lie algebras  corresponding to the Weyl symmetric ordering \cite{Meljanac-3} \cite{Meljanac-4} we express the generators $\tilde{M}_{AB}$ as
\begin{equation}\label{5.1.3}
\tilde{M}_{AB}=X_{CD}\tensor{\left( \frac{\mathbf{\tilde{C}}}{1-e^{-\mathbf{\tilde{C}}}}\right) }{_A_B^C^D},
\end{equation}
where
\begin{equation}\label{5.1.4} \tensor{\mathbf{\tilde{C}}}{_A_B^C^D}=-\frac{1}{2}\tensor{\tilde{C}}{_A_B_,_E_F^C^D}K^{EF}
\end{equation}
 and
\begin{equation}\label{5.1.5}
\tensor{\tilde{C}}{_A_B_,_E_F^C^D}=\frac{1}{2}\left[ -g_{BE}\left( \tensor{\delta}{_A^C}\tensor{\delta}{_F^D}-\tensor{\delta}{_A^D}\tensor{\delta}{_F^C}\right) +g_{AF}\left( \tensor{\delta}{_E^C}\tensor{\delta}{_B^D}-\tensor{\delta}{_E^D}\tensor{\delta}{_B^C}\right) -(A \leftrightarrow B)\right].
\end{equation}
Note that $\tilde{M}_{AB}=\left( SMS^{T}\right)_{AB} , \; X_{AB}=\left( SxS^{T}\right)_{AB}, \; K^{AB}=\left( S^{\ddag}k S^{-1}\right) ^{AB},\; $ where $S^{\ddag}=\left( S^{-1}\right) ^{T}.\;$
The Weyl realization of $\tilde{M}_{AB}$ in the terms of the generalized Heisenberg algebra in terms of $X_{AB}$ and $K^{AB}$ reads up to second order,
\begin{equation}\label{5.2.1}
\tilde{M}_{AB}=X_{AB}+\frac{1}{2}X_{CD}\tensor{\mathbf{\tilde{C}}}{_A_B^C^D}+\frac{1}{12}X_{CD}\tensor{\left( \mathbf{\tilde{C}}^{2}\right) }{_A_B^C^D},
\end{equation}
where $\tensor{\mathbf{\tilde{C}}}{_A_B^C^D}\;$ and $\;\tensor{\left( \mathbf{\tilde{C}}^{2}\right) }{_A_B^C^D}\;$ are as in \eqref{5.6} and \eqref{5.7} respectively, with $k \rightarrow K$. The indices are lowered and raised with the metric $g_{AB}$ and $g^{AB}$ respectively. Inserting $\mathbf{C}$ in \eqref{5.2.1} we find up to first order
\begin{equation}\label{5.17}
\tilde{M}_{AB}=X_{AB}+\frac{1}{2}\left( X_{AE}\tensor{K}{_B^E}- X_{BE}\tensor{K}{_A^E}\right)
\end{equation}
and
\begin{equation}
\left[ \tilde{M}_{AB}, K^{CD}\right] =i\left( \tensor{\delta}{_A^C}\tensor{\delta}{_B^D}-\tensor{\delta}{_A^D}\tensor{\delta}{_B^C}\right)+\frac{i}{2}\left( \tensor{\delta}{_A^C}\tensor{K}{_B^D}-\tensor{\delta}{_B^C}\tensor{K}{_A^D}+\tensor{\delta}{_B^D}\tensor{K}{_A^C}-\tensor{\delta}{_A^D}\tensor{K}{_B^C}\right).
\end{equation}

The Weyl realization $\tilde{M}^{AB}$ given in \eqref{5.1.3}-\eqref{5.1.5} enjoys the property
\begin{equation}
e^{\frac{1}{2}t^{AB}\tilde{M}^{AB}} \rhd 1=e^{\frac{1}{2}t^{AB}X^{AB}},
\end{equation}
where $t_{AB}$ are real numbers transforming as tensors, and action $\rhd$ is defined as
\begin{equation}
X_{AB} \rhd f(X_{EF})=X_{AB}f(X_{EF}), \quad K^{AB} \rhd f(X_{EF})=-i\frac{\partial f(X_{EF}) }{\partial X_{AB}}=\left[ K^{AB}, f(X_{EF})\right] .
\end{equation}
In particular
\begin{equation}
X_{AB} \rhd 1= X_{AB}, \; K_{AB} \rhd 1=0,
\end{equation}
\begin{equation}
K^{AB} \rhd e^{\frac{1}{2}t^{EF}X^{EF}}=t^{AB}e^{\frac{1}{2}t^{EF}X^{EF}}.
\end{equation}
Also, as in the previous subsection, we can write the generators $X_{AB}$ and $K^{AB}$ in terms of the four dimensional variables
\begin{align}
& K^{\mu 4}=\frac{Q^{\mu}}{M},\quad K^{\mu 5}=\frac{Y^{\mu}}{R}, \quad K^{45}=\frac{W}{MR},\\
&X_{\mu 4}=M X_{\mu}, \quad X_{\mu 5}=R P_{\mu}, \quad X_{45}=MRH,
\end{align}
so that
\begin{equation}
\left[ X_{\mu},Q_{\nu}\right] =i\delta_{\mu}^{\nu}, \quad \left[ P_{\mu},Y_{\nu}\right] =i\delta_{\mu}^{\nu}, \quad \left[ H,W\right] =i.
\end{equation}
\bigskip
\subsection{Weyl realization for dual $\kappa$ Minkowski spaces}
 The generators $\tilde{M}_{AB}$ can be written in terms of the four dimensional variables with Greek indices ($\tilde{M}_{\mu\nu},\; \tilde{X}_{\mu},\; \tilde{P}_{\mu},\; \tilde{H}$, expressed in terms of $M_{\mu\nu},\; X_{\mu}, \; P_{\mu},\;H,\;$ and $K_{\mu\nu},\; Q_{\mu},\;Y_{\mu},\; W$). They are
\begin{align}
&\tilde{M}_{\mu\nu}=X_{\mu\nu}+\frac{1}{2}\Big( X_{\mu\alpha}\left( \tensor{K}{_\nu^\alpha}-\frac{1}{M}g_{\nu 4}Q^{\alpha}-\frac{1}{R}g_{\nu 5}Y^{\alpha}\right)\notag\\
&+X_{\mu}\left( Q_{\nu}-\frac{1}{R}g_{\nu 5}W\right) +P_{\mu}\left( Y_{\nu}+\frac{1}{M}g_{\nu 4}W\right) -\left( u\leftrightarrow v\right) \Big),
\end{align}
\begin{align}
&\tilde{X}_{\mu}=X_{\mu}+\frac{1}{2}\Big(-\frac{1}{M}X_{\mu\nu}\left( g_{\alpha 4}K^{\nu\alpha}+\frac{1}{M}g_{44}Q^{\nu}+\frac{1}{R}g_{45}Y^{\nu}\right)
+\frac{1}{M}X_{\mu}\left( g_{\nu 4}Q^{\nu}-\frac{1}{R}g_{45}W\right)\notag \\
 &+X_{\nu}\left( \tensor{K}{_\mu^\nu}-\frac{1}{M}g_{\mu 4}Q^{\nu}-\frac{1}{R}g_{\mu 5}Y^{\nu}\right)
+\frac{1}{M}P_{\mu}\left( g_{\nu 4}Y^{\nu}+\frac{1}{M}g_{44}W\right) -H\left( Y_{\mu}+\frac{1}{M}g_{\mu 4}W\right) \Big),
\end{align}
\begin{align}
&\tilde{P}_{\mu}=P_{\mu}+\frac{1}{2}\Big(-\frac{1}{R}X_{\mu\nu}\left( g_{\alpha 5}K^{\nu\alpha}+\frac{1}{M}g_{45}Q^{\nu}+\frac{1}{R}g_{55}Y^{\nu}\right)
+\frac{1}{R}P_{\mu}\left( g_{\nu 5}Y^{\nu}+\frac{1}{M}g_{45}W\right)\notag \\
&+P_{\nu}\left( \tensor{K}{_\mu^\nu}-\frac{1}{R}g_{\mu 5}Y^{\nu}-\frac{1}{M}g_{\mu 4}Q^{\nu}\right)
+\frac{1}{R}X_{\mu}\left( g_{\nu 5}Q^{\nu}-\frac{1}{R}g_{55}W\right) +H\left( Q_{\mu}-\frac{1}{R}g_{\mu 5}W\right) \Big),
\end{align}
\begin{align}
&\tilde{H}=H+\frac{1}{2}\Big(\frac{1}{R}X_{\mu}\left( g_{\nu 5 }K^{\mu\nu}+\frac{1}{M}g_{45}Q^{\mu}+\frac{1}{R}g_{55}Y^{\mu}\right)\notag \\
&-\frac{1}{M}P_{\mu}\left( \frac{1}{M}g_{4 4}Q^{\mu}+\frac{1}{R}g_{45}Y^{\mu}+g_{\nu 4 }K^{\mu\nu}\right)
 +H\left(\frac{1}{R}g_{\mu 5} Y^{\mu}+\frac{1}{M}g_{\mu 4}Q^{\mu}\right) \Big).
\end{align}
For the dual $\kappa$-Minkowski spaces $g_{44}=0,\; g_{55}=0,\;$ see \eqref{4.21}-\eqref{4.30}.
The other realizations can be obtained from the Weyl realizations of $\tilde{M}_{AB}^{W}$ by using similarity transformation
\begin{equation}
\tilde{M}_{AB}=\mathcal{S}\tilde{M}_{AB}^{W}\mathcal{S}^{-1},
\end{equation}
where $\mathcal{S}=\exp(G)$, with $G=XF(K)$. The corresponding realizations will be linear in $X$ and can be written as a series in $K$. The corresponding coproducts, star products and twists can be obtained by using this similarity transformation \cite{Meljanac-1}.

\section{Coproduct, star product and twist in Weyl realization}
\label{s-6}
Formulae for coproduct and deformed addition of momenta for the generalized Yang model and dual $\kappa$-Minkowski spaces are written in terms of the curved metric $g_{AB}$ and are deduced using results of \cite{MM-4} \cite{Lukierski-3} \cite{Meljanac-4}. The corresponding coproducts for the momenta, $\Delta K^{\mu\nu},\; \Delta Q^{\mu}, \;  \Delta Y^{\mu},\; \Delta W$ are coassociative, see \cite{Battisti-1}. Defining
\begin{equation}
\exp \Big( \frac{i}{2}s^{AB}\tilde{M}_{AB}\Big) \exp\Big( \frac{i}{2}t^{CD}\tilde{M}_{CD}\Big)=\exp \Big( \frac{i}{2}\left(  s^{AB}\oplus t^{AB}\right)  \tilde{M}_{AB}\Big)=\exp\Big( \frac{i}{2}\mathcal{D}^{AB}(s,t)\tilde{M}_{AB}\Big),
\end{equation}
where $s^{AB}$ and $t^{AB}$ transform as tensors of corresponding orthogonal algebra. One has at first order
\begin{equation}\label{6.2}
\mathcal{D}^{AB}\left( s^{CD}, t^{CD}\right) =s^{AB}+t^{AB}-\frac{1}{2}\left( s^{AC}\tensor{t}{^B_C}-s^{BC}\tensor{t}{^A_{C}}\right).
\end{equation}
In the following all formulae are written in the first order. The coproduct $\Delta K^{AB}$ is
\begin{equation}
\Delta K^{AB}=\mathcal{D}^{AB}\left(  K^{AB}\otimes 1, 1 \otimes  K^{AB}\right)  =\Delta_{0} K^{AB}-\frac{1}{2}\left(  K^{AC}\otimes \tensor{K}{^B_C}-K^{BC}\otimes \tensor{K}{^A_C}\right),
\end{equation}
where $\Delta_{0} K^{AB}=K^{AB}\otimes 1+1 \otimes  K^{AB}$. In components, it reads
\begin{align}
\Delta K^{\mu\nu}&=\Delta_{0} K^{\mu\nu}-\frac{1}{2}\Big( K^{\mu\alpha}\otimes \tensor{K}{^\nu_\alpha} +\frac{1}{M^{2}} g_{44} Q^{\mu}  \otimes Q^{\nu} +  \frac{1}{R^{2}} g_{55} Y^{\mu}  \otimes Y^{\nu}\notag \\
&+\frac{1}{MR} g_{45} \left( Q^{\mu}  \otimes Y^{\nu}+Y^{\mu}  \otimes Q^{\nu}\right)+ \frac{1}{M} g_{\alpha 4} \left( K^{\mu \alpha}  \otimes Q^{\nu}+Q^{\mu}  \otimes K^{\nu\alpha}\right) \notag \\
& + \frac{1}{R} g_{\alpha 5} \left( K^{\mu \alpha}  \otimes Y^{\nu}+Y^{\mu}  \otimes K^{\nu\alpha}\right)-\left( u\leftrightarrow v\right)       \Big),
\end{align}
\begin{align}
\Delta Q ^{\mu}&=\Delta_{0}Q^{\mu}-\frac{1}{2}\Big(-K^{\mu\alpha}  \otimes Q_{\alpha}  +Q_{\alpha} \otimes K^{\mu\alpha} +\frac{1}{MR}g_{45} \left( Q^{\mu}\otimes W -W \otimes Q^{\mu}  \right) \notag \\
&+\frac{1}{R^{2}}g_{55} \left( Y^{\mu}\otimes W -W \otimes Y^{\mu}  \right)+\frac{1}{R}g_{\alpha 5} \left( K^{\mu \alpha}\otimes W -W \otimes K^{\mu \alpha}  \right)\notag \\
&-\frac{1}{M}g_{\alpha 4} \left( Q^{\mu}\otimes Q^{\alpha} -Q^{\alpha} \otimes Q^{\mu}  \right) -\frac{1}{R}g_{\alpha 5} \left( Y^{\mu}\otimes Q^{\alpha} -Q^{\alpha} \otimes Y^{\mu}  \right)\Big),
\end{align}
\begin{align}
\Delta Y ^{\mu}&=\Delta_{0}Y^{\mu}-\frac{1}{2}\Big(-K^{\mu\alpha}  \otimes Y_{\alpha}  +Y_{\alpha} \otimes K^{\mu\alpha} -\frac{1}{MR}g_{45} \left( Y^{\mu}\otimes W -W \otimes Y^{\mu}  \right) \notag \\
&+\frac{1}{M^{2}}g_{44} \left(W\otimes Q^{\mu} - Q^{\mu}\otimes W  \right)-\frac{1}{M}g_{\alpha 4} \left( K^{\mu \alpha}\otimes W -W \otimes K^{\mu \alpha}  \right)\notag \\
&-\frac{1}{M}g_{\alpha 4} \left( Q^{\mu}\otimes Y^{\alpha} -Y^{\alpha} \otimes Q^{\mu}  \right) -\frac{1}{R}g_{\alpha 5} \left( Y^{\mu}\otimes Y^{\alpha} -Y^{\alpha} \otimes Y^{\mu}  \right)\Big),
\end{align}
\begin{align}
\Delta W &= \Delta_{0}W-\frac{1}{2}\Big( Q^{\alpha}\otimes Y_{\alpha}-Y^{\alpha}\otimes Q_{\alpha}+\frac{1}{R}g_{\alpha 5} \left( Y^{\alpha}\otimes W -W \otimes Y^{\alpha}  \right) \notag \\
&+\frac{1}{M}g_{\alpha 4} \left( Q^{\alpha}\otimes W -W \otimes Q^{\alpha}  \right)\Big).
\end{align}
The antipodes of $\;K^{\mu\nu}, Q^{\mu}, Y^{\mu}\;$ and $\;W\;$ are trivial. The star product is defined as
\begin{equation}
e^{\frac{1}{2}s^{AB}X_{AB}}\star e^{\frac{1}{2}t^{CD}X_{CD}}=e^{\frac{1}{2}\mathcal{D}^{AB}(s,t)X_{AB}},
\end{equation}
where $\mathcal{D}^{AB}(s,t)$ is given in \eqref{6.2}. This star product is associative.

Defining $\;\mathcal{D}^{\mu}=\mathcal{D}^{\mu 4},\; \bar{\mathcal{D}}^{\mu}=\mathcal{D}^{\mu 5},\; \mathcal{D}=\mathcal{D}^{4 5},\;$ the four dimensional expression of $\mathcal{D}^{AB}(s,t)\;$ is given by
\begin{align}
\mathcal{D}^{\mu\nu}(s,t)&=s^{\mu\nu}+t^{\mu\nu}-\frac{1}{2}\Big(s^{\mu\rho}\tensor{t}{^\nu_\rho}+\frac{1}{M^{2}}g_{44}s^{\mu}t^{\nu}+  \frac{1}{R^{2}}g_{55}\bar{s}^{\mu}\bar{t}^{\nu} +\frac{1}{MR}g_{45}\left(s^{\mu}\bar{t}^{\nu} +\bar{s}^{\mu}t^{\nu} \right) \notag \\
&+ \frac{1}{M}g_{\mu 4}\left(s^{\mu \rho}t^{\nu} +s^{\mu}t^{\nu\rho} \right) +\frac{1}{R}g_{\mu 5}\left(s^{\mu \rho}\bar{t}^{\nu} +\bar{s}^{\mu}t^{\nu\rho}\right) -\left( u\leftrightarrow v\right)    \Big),
\end{align}
\begin{align}
\mathcal{D}^{\mu}(s,t)&=s^{\mu}+t^{\mu}-\frac{1}{2}\Big(s^{\mu\rho}\tensor{t}{_\rho}-t^{\mu\rho}\tensor{s}{_\rho} + \frac{1}{MR}g_{45}\left(s^{\mu}t -st^{\mu} \right) \notag \\
&+ \frac{1}{R^{2}}g_{55}\left(\bar{s}^{\mu}t -s\bar{t}^{\mu} \right) + \frac{1}{R}g_{\mu 5}\left(s^{\mu \rho}t -st^{\mu\rho} \right)   \Big),
\end{align}
\begin{align}
\bar{\mathcal{D}}^{\mu}(s,t)&=\bar{s}^{\mu}+\bar{t}^{\mu}-\frac{1}{2}\Big(s^{\mu\rho}\tensor{\bar{t}}{_\rho}-\bar{s}_{\rho}t^{\mu\rho}-\frac{1}{MR}g_{45}\left(\bar{s}^{\mu}t -s\bar{t}^{\mu} \right) \notag \\
&+ \frac{1}{M^{2}}g_{44}\left(s^{\mu}t -st^{\mu}\right) -\frac{1}{M}g_{\mu 4}\left(s^{\mu \rho}t -st^{\mu\rho} \right)  \Big),
\end{align}
\begin{equation}
\mathcal{D}(s,t)=s+t-\frac{1}{2}\left( s^{\rho}\bar{t}_{\rho}-\bar{s}^{\rho}t_{\rho}\right),
\end{equation}
where we have defined the components of tensors $t^{AB}$ corresponding to the orthogonal algebra as $\; t^{\mu}=t^{\mu 4},\; \bar{t}^{\mu}=t^{\mu 5},\; t=t^{45}\;$ and analogously for $s^{AB}$.\\
\bigskip

Now, we construct the twist operator at first order as in \cite{MM-4} \cite{Meljanac-4}. The twist is defined as a bilinear operator such that
\begin{equation}
\Delta K^{AB}=\mathcal{F}\left( \Delta_{0} K^{AB}\right) \mathcal{F}^{-1}
\end{equation}
for momenta $K^{AB}$ belonging to corresponding orthogonal algebra. The twist was introduced in the context of the NC geometry in \cite{Chaichain} \cite{Wess} as a useful tool in construction of quantum field theories. In a Hopf algebroid approach \cite{Meljanac-5} \cite{juric-1} \cite{juric-2} the twist can be written as
\begin{equation}
\mathcal{F}^{-1}\equiv e^{F}=e^{-\frac{i}{2}K^{AB}\otimes X_{AB}}e^{\frac{i}{2}K^{CD}\otimes \tilde{M}_{CD}}.
\end{equation}
Using the BCH formula one gets
\begin{equation}\label{6.15}
F=\frac{i}{2}K^{AB}\otimes \left( \tilde{M}_{AB}-X_{AB}\right)
\end{equation}
and substituing \eqref{5.17} into  \eqref{6.15} one has
\begin{equation}
F=\frac{i}{2}K^{AC}\otimes X_{AB}\tensor{K}{_{C}^{B}}.
\end{equation}
In terms of components, one can write
\begin{align}
F&=\frac{i}{2}\Big( K^{\mu\nu}\otimes \left[X_{\mu\alpha}\left( \tensor{K}{_{\nu}^{\alpha}}-\frac{1}{M} g_{\nu 4}Q^{\alpha}-\frac{1}{R} g_{\nu 5}Y^{\alpha}\right) +X_{\mu}\left( Q_{\nu}-\frac{1}{R} g_{\nu 5}W\right) +P_{\mu} \left( Y_{\nu}+\frac{1}{M} g_{\nu 4}W\right) \right] \notag \\
&+Q^{\mu}\otimes \Big[ -\frac{1}{M} X_{\mu\nu}\left( g_{\alpha 4}K^{\nu\alpha}+\frac{1}{M} g_{44}Q^{\nu}+\frac{1}{R} g_{45}Y^{\nu}\right) +\frac{1}{M} X_{\mu}\left( g_{\alpha 4}Q^{\alpha}-\frac{1}{R} g_{45}W\right)\notag \\
&+\frac{1}{M} P_{\mu}\left( g_{\alpha 4}Y^{\alpha}+\frac{1}{M} g_{44}W\right) \Big] +X_{\nu}\left( \tensor{K}{_\mu^\nu}-\frac{1}{M} g_{\mu 4} Q^{\nu}-\frac{1}{R} g_{\nu 5}Y^{\nu}\right) +H\left( Y_{\mu}+\frac{1}{M} g_{\mu 4} W\right)\Big]\notag \\
&+P^{\mu}\otimes \Big[ -\frac{1} {R}X_{\mu\nu}\left( g_{\alpha 5}K^{\nu\alpha}+\frac{1}{M} g_{45}Q^{\nu}+\frac{1}{R} g_{55}Y^{\nu}\right) +\frac{1}{R} X_{\mu}\left( g_{\alpha 5}Q^{\alpha}-\frac{1}{R} g_{55}W\right)\notag \\
&+\frac{1}{R} P_{\mu}\left( g_{\alpha 5}Y^{\alpha}+\frac{1}{M} g_{45}W\right) \Big] +P_{\nu}\left( \tensor{K}{_\mu^\nu}-\frac{1}{M} g_{\mu 4} Q^{\nu}-\frac{1}{R} g_{\nu 5}Y^{\nu}\right) +H\left( Q_{\mu}-\frac{1}{R} g_{\mu 5} W\right)\Big]\notag \\
&+ W \otimes \Big[ \frac{1}{R} X_{\mu}\left( g_{\nu 5}K^{\mu\nu}+\frac{1}{M} g_{45}Q^{\mu}+\frac{1}{R} g_{55}Y^{\mu}\right)-\frac{1} {M}P_{\mu}\left( g_{\nu 4}K^{\mu\nu}+\frac{1}{M} g_{44}Q^{\mu}+\frac{1} {R}g_{45}Y^{\mu}\right) \notag \\
&+H\left( \frac{1} {R}g_{\mu 5}Y^{\mu}+\frac{1}{M} g_{\mu 4}Q^{\mu}\right) \Big]\Big).
\end{align}
The coproducts, star product and twist for the Yang model for different orthogonal algebras can be derived from the above results setting $u=v=1,\; \varphi=\psi =0\;$ and $a_{\mu}=b_{\mu}=0$.

\section{Reduced dual $\kappa$-Minkowski spaces and $\kappa$- Poincar\'{e} algebras}
\label{s-7}
Let us now consider the reduced Yang models, introduced in \cite{MM-5}. If the quadratic Casimir operator of the orthogonal algebra \eqref{2}, $\frac{1}{2}M_{AB} M^{AB}$ with metric $\eta_{AB}$, is restricted to  a constant value $\epsilon_{1}\epsilon_{2}M^2R^2$, then the generator $M_{45}$ can be expressed in terms of the other 14 generators. Consequently, for a Yang algebra \eqref{0.1}-\eqref{0.5}, one can define a nonlinear algebra with 14 generators, called reduced Yang algebra, with
\begin{equation}\label{7.1}
\hat h=\sqrt{1-{\epsilon_2\over R^2}\hat x_{\mu} \hat x^{\mu}-{\epsilon_1\over M^2}\hat p_{\mu}\hat p^{\mu}-{\epsilon_1\epsilon_2\over2M^2R^2}M_{\alpha\beta}M^{\alpha\beta}}.
\end{equation}
It has been shown in \cite{MM-5}, that this expression for $\hat h$ satisfies the relations (\ref{0.3}), (\ref{0.4}) and hence all the Yang algebra commutation relations \eqref{0.1}-\eqref{0.5} hold.

Following the construction in sections \ref{s-2}, \ref{s-3} and \ref{s-4}, we can define $\tilde H$, $\tilde X_\mu$, $\tilde P_\mu$ and $M_{\mu\nu}$, using the above expression of $\hat h$ \eqref{7.1}. For the dual $\kappa$-Minkowski space and $\kappa$ -Poincar\'{e} algebra, using the results of subsection \ref{sub-4.2} we obtain the relations for $\tilde H$, $\tilde X_\mu$, $\tilde P_\mu$, $M_{\mu\nu}$ in the terms of  $\hat x_\mu$, $\hat p_\mu$, $\hat{h}$, $M_{\mu\nu}$. The relation \eqref{4.29} for $\tilde H$  is the same but with $\hat h$ \eqref{7.1} expressed in the terms of $\tilde X_\mu$, $\tilde P_\mu$, $M_{\mu\nu}$.

In the special case $u=v=1$, $\phi=\psi=0$,
\begin{equation}
\tilde X_\mu=\hat x_\mu+{1\over\kappa}a_\rho M_{\mu\rho}\qquad\tilde P_\mu=\hat p_\mu+{1\over\tilde\kappa}b_\rho M_{\mu\rho}\qquad
\end {equation}
\begin{equation}
\tilde H=\hat h+{1\over\kappa}a\cdot\tilde P-{1\over\tilde\kappa}b\cdot\tilde X-{1\over\kappa\tilde\kappa}a_\rho b_\sigma M_{\rho\sigma},
\end{equation}
taking into account that in \eqref{7.1} the following changes should be made
\begin{equation}
-\frac{\epsilon_{1}}{M^{2}}=\frac{a^{2}}{\kappa^{2}},\; -\frac{\epsilon_{2}}{R^{2}}=\frac{b^{2}}{\tilde{\kappa}^{2}}.
\end{equation}
Relations \eqref{4.26}-\eqref{4.28} are still valid.
Moreover, in the limit $\tilde\kappa\to\infty$ we obtain
\begin{equation}
\tilde X_\mu=\hat x_\mu+{1\over\kappa}a_\rho M_{\mu\rho}\qquad\tilde P_\mu=\hat p_\mu,
\end{equation}
\begin{equation}
\tilde{H}=\sqrt{1+{a^{2}\over\kappa^2}\tilde P_{\mu}\tilde P^{\mu}}+{1\over\kappa}a\cdot\tilde P
\end{equation}
with the properties $[\tilde H,\tilde X_\mu]=-{1\over\kappa}a_\mu\tilde H$ and $[\tilde H,\tilde P_\mu]=0$.

The generator $\tilde H$ is the shift operator  and the case $\tilde\kappa\to\infty$ corresponds to the natural realization of $\kappa$-Poincar\'{e} algebra \cite{MK}.

\section{Concluding remarks}

The Yang model describes a noncommutative geometry defined on a curved  background. Its interest for physics resides in possible applications to  quantum cosmology.
From a mathematical point of view, its most remarkable property is the existence of a Born duality between positions and momenta.

The model is based on the Yang algebra \eqref{0.1} – \eqref{0.5}, which depends on a mass $M$, a length $R$ and the discrete parameters $\epsilon_{1} =\pm 1,\; \epsilon_{2} =\pm 1$, and is isomorphic  either to $o(1,5),\; o(2,4)$ or $o(3,3)$ algebras with flat metrics $\eta_{AB}$, depending on the values of $\epsilon_{1}$ and $\epsilon_{2}$.

Performing general linear transformations that keep Lorentz generators unchanged, we have constructed a class of algebras depending on the parameters $M, R, \kappa,  \tilde {\kappa}$, vectors $a_{\mu}, b_{\mu}$ and scalars $u, v, \epsilon_{1} \varphi+ \epsilon_{2} \psi$, which are isomorphic either to $o(1,5,g), o(2,4,g)$ or $o(3,3,g)$ algebras with metric $g_{AB}$ \eqref{4.20} satisfying $g_{\mu \nu} = \eta_{\mu \nu}$. This construction unifies all Lorentz invariant models isomorphic to orthogonal algebras and generalizes the results of \cite{MMM}.

In particular, the dual $\kappa$-Minkowski and $\kappa$-Poincar\'{e} algebras introduced in \cite{LMP2} are obtained demanding $g_{44} = 0,\; g_{55} = 0$, with $\det g \neq 0$, leading to the relations \eqref{4.30}, $\frac{a^{2}}{\kappa^{2}}=-\epsilon_{1}\frac{u^{2}}{M^{2}}$ and $\frac{b^{2}}{\tilde{\kappa}^{2} }= -\epsilon_{2} \frac{v^{2}}{R^{2}}$. The interest of such models is that they extend the standard $\kappa$-Minkowski space
to a dual formulation independent from the Snyder framework.
In the limit $M \rightarrow \infty\:$ the vector $a_{\mu}$ becomes light-like,  $a^{2} = 0$, while for $R \rightarrow \infty$ the vector $b_{\mu}$ becomes light-like, $b^{2} = 0$.  If in addition, $\kappa \rightarrow \infty,\; \tilde{\kappa} \rightarrow \infty$, the algebra reduces to the Heisenberg algebra.

We have also obtained the Weyl realization of dual $\kappa$-Minkowski and $\kappa$-Poincar\'{e} algebras in terms of the metric $g$ and thence the corresponding coproduct, star product and twist. Finally, following the suggestion
 of \cite{MM-5}, we have constructed reduced $\kappa$-Minkowski spaces and $\kappa$-Poincar\'{e} algebras, where the generator $\tilde{H}$ is no longer taken as independent, but as a function of the other 14 generators of the algebra.

In this paper, we have considered only formal aspects of the theory. Possible physical effects can arise in the nonrelativistic limit, when one may investigate the generalization of uncertainty relations and of dynamics coming from the deformation of the Heisenberg commutation relations, as in \cite{MM-6,MM-5}. One may also apply our results to models of doubly special relativity with a nontrivial spacetime background.

\subsection*{Acknowledgement}
We thank Jerzy Lukierski for useful discussions.
S. Mignemi acknowledges contribution of Gruppo Nazionale di Fisica Matematica and of COST Action CA23130.


\begin{thebibliography}{99}

\bibitem{gup} M. Maggiore, A Generalized uncertainty principle in quantum gravity, \PL{B 304} 63 (1993).
M. Maggiore, The Algebraic structure of the generalized uncertainty principle, \PL{B 319} 83 (1993).
A. Kempf, G. Mangano, R.B. Mann, Hilbert space representation of the minimal length uncertainty relation, \PR{D 52} 1108 (1995).

\bibitem{DF}  S. Doplicher, K. Fredenhagen, J. E. Roberts, {The quantum structure of spacetime at the Planck scale and quantum fields,} \CMP{172} 187 (1995).

\bibitem{KG} J. Kowalski-Glikman, De Sitter space as an arena for doubly special relativity, \PL{B 547} 291 (2002).

\bibitem{Ya} C.N. Yang, {On Quantized Space-Time}, \PR{72} 874 (1947).

\bibitem{Sn} H.S. Snyder, {Quantized Space-Time},  \PR{71} 38 (1947).

\bibitem{Bo} M. Born, A suggestion for unifying quantum theory and relativity, \PRS{A 165} 291 (1938).

\bibitem{KL} V.V. Khruschev, A.N. Leznov, {Relativistic invariant Lie algebras for kinematical observables in quantum space time,} {\textit{Grav. Cosmol.}} {\bf 9} 159 (2003).

\bibitem{Meljanac-1}
\bibp{S. Meljanac, R. \v{S}trajn}{Deformed Quantum Phase Spaces, Realizations, Star Products and Twists} {\textit{SIGMA}}{ \textbf{18} 022 (2022)}.

\bibitem{Kowalski-1}
\bibp{J. Kowalski-Glikman, L. Smolin}{Triply Special Relativity}{\PR{D 70}}{ 065020 (2004)}.

\bibitem{Mignemi-1}
\bibp{S. Mignemi}{The Snyder-de Sitter model from six dimensions}\CQG{26} 245020 (2009).

\bibitem{MM-1}  S. Meljanac, S. Mignemi, {Generalizations of Snyder model to curved spaces,} \PL{B 833} 137289 (2022).

\bibitem{MM-2}
\bibp{S. Meljanac, S. Mignemi}{Noncommutative Yang model and its generalizations} {\JMP{64}}  {023505 (2023)}.

\bibitem{Lukierski-1}
\bibp{ J. Lukierski, S.  Meljanac, S.  Mignemi, A. Pacho\l{}}{Quantum pertubative solutions of extended Snyder and Yang models with spontaneous symmetry breaking}{\PL{B 847}} {138261 (2023)}.

\bibitem{Battisti-1}
\bibp{M.V. Battisti, S. Meljanac}{Scalar Field Theory on Non-commutative Snyder Space-Time}{\PR{D 82}} {024028 (2010)}.

\bibitem{MM-3}
\bibp{S. Meljanac, S. Mignemi}{Associative realizations of the extended Snyder model}{\PR{D 102}} {126011 (2020)}.

\bibitem{Girelli}
\bibp{F. Girelli, E.R. Livine}{Scalar field theory in Snyder space-time: alternatives}{\JHEP{1103}} {132 (2011)}.


\bibitem{Guo}
\bibp{H.G. Guo, C.G. Huang, H.T. Wu}{Yang's Model as Triply Special Relativity and the Snyder's Model--de Sitter Special Relativity Duality}{\PL{B 663}} {270 (2008)}.

\bibitem{Banerjee}
\bibp{R. Banerjee, K. Kumar, D. Roychowdhury}{Symmetries of Snyder-de Sitter space and relativistic particle dynamics}{\JHEP{03}} {060 (2011)}.

\bibitem{Lukierski-2}
\bibp{J. Lukierski, M. Woronowicz}{Spinorial Snyder and Yang models from superalgebras and noncommutative quantum superspaces}{\PL{B 824}} {136783 (2021)}.

\bibitem{Mignemi-2}
\bibp{S. Mignemi}{Doubly special relativity in de Sitter spacetime}{\textit{Annalen Phys.}}{ \textbf{522} 924 (2010)}.

\bibitem{Buric}
\bibp{M. Buri\'{c}, M. Wohlgennant}{Geometry of the Grosse-Wulkenhaar Model}{\JHEP{1003}} {053 (2010)}.

\bibitem{Ballesteros}
\bibp{A. Ballesteros, G. Gubitosi, F. Mercati}{Interplay between spacetime curvature, speed of light and quantum deformations of relativistic symmetries} {\textit{Symmetry}} {\textbf{13} 2099 (2021)}.

\bibitem{Aschieri}
\bibp{P. Aschieri, A. Borowiec, A. Pacho\l}{Dispersion relations in $\kappa$-noncommutative cosmology} {\JCAP{04}} {025 (2021)}.

\bibitem{others}J.J. Heckman, H. Verlinde, Covariant Non-Commutative Space-Time, \NP{B 894} 58 (2015).
P. Nandi, F.G. Scholtz, `The hidden Lorentz covariance of quantum mechanics, \textit{Annals Phys.} \textbf{464} 169643 (2024).
D. Roumelioti, S. Stefas, G. Zoupanos, Fuzzy Gravity: Four-Dimensional Gravity on a Covariant Noncommutative Space and Unification with Internal Interactions, \arx{2407.07044}.
A. Manta, H.C. Steinacker, Minimal covariant quantum space-time, \arx{2502.02498}.

\bibitem{MS-1}
\bibp{S. Meljanac, D. Meljanac, A. Samsarov, M. Stoji\'c}{Kappa-deformed Snyder space} {\textit{Mod. Phys. Lett.}} {\textbf{A 25} 579 (2010)}.

\bibitem{MS-2}
\bibp{S. Meljanac, D. Meljanac, A. Samsarov, M. Stoji\'c}{Kappa Snyder deformations of Minkowski spacetime, realizations and Hopf algebra} {\PR{D 83}} {065 009  (2011)}.

\bibitem{MM-4}
\bibp{S. Meljanac, S. Mignemi}{Associative realizations of $\kappa$-deformed extended Snyder model}{\PR{D 104}} {086006 (2021)}.

\bibitem{Lukierski-3}
\bibp{ J. Lukierski, S.  Meljanac, S.  Mignemi, A. Pacho\l{}}{Generalized quantum phase spaces for the $\kappa$-deformed extended Snyder model}{\PL{B 838}} {137709 (2023)}.

\bibitem{Meljanac-3}
\bibp{S. Meljanac, T. Martini\'{c} Bila\'{c},  S. Kre\v{s}i\'{c}--Juri\'{c}}{Generalised Heisenberg algebra applied to realizations of the orthogonal, Lorentz and Poincar\'{e} algebras and their dual extensions}{\JMP{61}} {051705 (2020)}.

\bibitem{MMM}  T. Martini\'c-Bila\'c,  S. Meljanac, S. Mignemi, {Realizations and star-product of doubly $\kappa$-deformed Yang models,} \EPJ{C 84} 846 (2024).

\bibitem{LMP1} \bibp{J. Lukierski, S. Meljanac, S. Mignemi, A. Pacho\l\ and M. Woronowicz} {From Snyder space-times to doubly $\kappa$-dependent Yang quantum phase spaces and their generalizations} {\PL{B 854} 138729 (2024)}.

\bibitem{kappa}  J. Lukierski, H. Ruegg, A. Novicki, V.N. Tolstoi, {q-deformation of Poincar\'e algebra,} \PL{B 264} 331 (1991).
J. Lukierski and  H. Ruegg, {Quantum $\kappa$-Poincar\'e in Any Dimensions,} \PL{B 329} 189 (1994).

\bibitem{LMP2} J. Lukierski, S. Meljanac, S. Mignemi, A. Pacho\l, M. Woronowicz, {Towards new relativistic doubly $\kappa$-deformed D=4 quantum phase spaces,} \arx{2410.02339}.

\bibitem{MM-5}
\bibp{S. Meljanac, S. Mignemi}{Reduced Yang model and noncommutative geometry of curved spacetime} \arx{2503.23146}.

\bibitem{MK} D. Kova\v cevi\'c, S. Meljanac, {Kappa-Minkowski spacetime, Kappa-Poincaré Hopf algebra and realizations,} \JoP  {A 45} 135208 (2012).
S. Meljanac, S. Kre\v si\'c-Juri\'c, {Differential structure on kappa-Minkowski space, and kappa-Poincar\'{e} algebra} \IJMP{A 26} 3385 (2011).

\bibitem{Meljanac-4}
\bibp{S. Meljanac, Z. \v{S}koda, S. Kre\v{s}i\'{c}--Juri\'{c}} {Symmetric ordering and Weyl realizations for quantum Minkowski spaces} {\JMP{63}} {123508 (2022)}.

\bibitem{Chaichain}
\bibp{M. Chaichain, P. Kulish, K. Nishijima, A.Tureanu}{On a Lorentz-Invariant Interpretation of Noncommutative Space-Time and Its Implications on Noncommutative QFT}{\PL{B 604}} {98 (2004)}.

\bibitem{Wess}
\bibp{J. Wess}{in Mathematical, Theoretical and Phenomenological Chalenges Beyond the Standard Model ed. by G. Djor\dj evi\'c, L. Ne\v{s}i\'{c} and J.Wess} {World Scientific, 2005.} {arXiv:hep-th/0408080}.

\bibitem{Meljanac-5}
\bibp{ S. Meljanac, D. Meljanac, S. Mignemi, R. \v{S}trajn} {Snyder--type spaces, twisted Poincar\'{e} algebra and addition of momenta}
{\IJMP{A 32}} {1750172 (2017)}.

\bibitem{juric-1}
\bibp{T. Juri\'c, S. Meljanac, R. \v{S}trajn} {$\kappa$-{P}oincar\'e--{H}opf algebra and {H}opf algebroid structure of phase space from twist} {\PL{A 377}} {2472--2476 (2013)}.

\bibitem{juric-2}
\bibp{T. Juri\'c, D. Kova\v{c}evi\'c, S. Meljanac} {$\kappa$-deformed phase space, {H}opf algebroid and twisting} {\textit{SIGMA}} {\textbf{10} 106 (2014)}.

\bibitem{MM-6} S. Meljanac, S. Mignemi, Quantum mechanics of the nonrelativistic Yang model, \arx{2411.06443}.

\end{thebibliography}
\end{document}